\documentclass{aa}  
\usepackage{verbatim}
\usepackage{booktabs, multirow}
\usepackage{lipsum} 
\usepackage{graphicx}
\usepackage{txfonts}
\usepackage{amssymb}
\usepackage{xcolor}
\usepackage{natbib}
\usepackage[colorlinks=true,citecolor=blue]{hyperref}
\usepackage{mathtools}
\defcitealias{Danforth2016}{D16}
\defcitealias{Prakash2017}{G17} 
\def\lya{Ly$\alpha$ } 
 
\def\h2{H$_2$} 
\def\hi{H{\sc i}}

\def\kms{km s$^{-1}$~}

\usepackage[normalem]{ulem}

\begin{document} 

\title{FLAME: Fitting \lya absorption lines using machine learning}
   \author{Priyanka Jalan
          \inst{1}
          \and
            Vikram Khaire
          \inst{2, 3}
            \and
         M. Vivek
          \inst{4}
         \and 
          Prakash Gaikwad
          \inst{5}
          }

   \institute{Center for Theoretical Physics of the Polish Academy of Sciences, Al. Lotnik\'{o}w 32/46, 02-668 Warsaw, Poland\\
              \email{priyajalan14@gmail.com}
              \and
              Indian Institute of Space Science and Technology, Thiruvananthapuram, Kerala 695547, India
              \and
              Physics Department, Broida Hall, University of California Santa Barbara, Santa Barbara, CA 93106-9530, USA
              \and
              Indian Institute of Astrophysics, Koramangala, Bengaluru, Karnataka 560034, India
              \and
             Max-Planck-Institut f\"{u}r Astronomie, K\"{o}nigstuhl 17, D-69117 Heidelberg, Germany}

   \date{}

\abstract
{
We introduce FLAME, a machine-learning algorithm designed to fit Voigt profiles to H~{\sc i} Lyman-alpha (\lya) absorption lines using deep convolutional neural networks. FLAME integrates two algorithms: the first determines the number of components required to fit \lya absorption lines, and the second calculates the Doppler parameter $b$, the H~{\sc i} column density N$_{\rm HI}$, and the velocity separation of individual components. For the current version of FLAME, we trained it on low-redshift \lya forests observed with the far-ultraviolet gratings of the Cosmic Origin Spectrograph (COS) on board the Hubble Space Telescope (HST). Using these data, we trained FLAME on $\sim$ $10^6$ simulated Voigt profiles ---which we forward-modeled to mimic \lya absorption lines observed with HST-COS--- in order to classify lines as either single or double components and then determine Voigt profile-fitting parameters. 
FLAME shows impressive accuracy on the simulated data, identifying more than 98\% (90\%) of single (double) component lines. It determines $b$ values within $\approx \pm{8}~(15)$ km s$^{-1}$ and log $N_{\rm HI}/ {\rm cm}^2$ values within $\approx \pm 0.3~(0.8)$ for 90\% of the single (double) component lines. However, when applied to real data, FLAME's component classification accuracy drops by $\sim$ 10\%. Nevertheless, there is reasonable agreement between the $b$ and N$_{\rm HI}$ distributions obtained from traditional Voigt profile-fitting methods and FLAME's predictions. Our mock HST-COS data analysis, designed to emulate real data parameters, demonstrates that FLAME is able to achieve consistent accuracy comparable to its performance with simulated data. This finding suggests that the drop in FLAME's accuracy when used on real data primarily arises from the difficulty in replicating the full complexity of real data in the training sample. In any case, FLAME's performance validates the use of machine learning for Voigt profile fitting, underscoring the significant potential of machine learning for detailed analysis of absorption lines.
}


\maketitle

\section{Introduction}
\label{sect-intro}
The gas that permeates the space between galaxies is called the intergalactic medium (IGM).  One of the best ways to explore the IGM is to study the large range of  \lya absorption lines present in quasar spectra known as the \lya forest \citep[][]{Rauch1998, Meiksin2009}. These absorption lines result from the quasar's continuum being absorbed by the redshifted \lya (1215.67 \AA) resonance line of the neutral hydrogen gas. The \lya forest has been shown to be an exceptional tool for studying the thermal state of the IGM \citep[e.g.,][]{Schaye2001a, Bolton2008, Lidz2011, Hiss2018, Gaikwad2021, Hu2023}, the intensity of the ionizing ultraviolet background \citep[e.g.,][]{Bolton2007, Becker2013, Gaikwad2016, Khaire2019, Hu2023}, and a wide range of cosmological parameters, including the mass of neutrinos \citep[][]{McDonald2006, Baur2017, Yeche2017} and dark matter properties \citep[][]{Busca2013, Viel2013, Irsic2017, Alam2021}.

The \lya forest at high redshift shows remarkable consistency with theoretical expectations for the IGM, such as the expected Gunn-Peterson troughs in $z\sim 6$ quasar spectra \citep[][]{Fan06, Bosman2018, Eilers2018} and the peak in temperature around the epoch of He~{\sc ii} reionization \citep[e.g.,][]{Walther19, Gaikwad2021}, which is believed to conclude around $z\sim 3$ \citep[][]{McQuinn09, Shull10, Worseck11, Khaire17sed}.

In contrast, the low-redshift ($z<1$) \lya forest has yielded several unexpected results, prompting new investigations. For instance, the distribution of line widths in the \lya forest at $z<0.5$ is broader than that reproduced by simulations of the IGM  \citep[][]{Viel17, Gaikwad2017} and there is evidence of a higher-than-expected temperature at $z\sim 1$ \citep[][]{Hu2023}. Furthermore, the epoch $z<1$ is critically important for galaxy formation, as it is during this period that feedback from galaxy formation \citep[][]{Springel05, Hopkins08, Bolton2017, Weinberger17, Dave2019} is believed to have a significant impact on the galaxies in order to explain the observed properties of galaxies and the sharp decline in star formation rate \citep[][]{Madau2014, Khaire2015a}. Additionally, this is the epoch where more than 30\% of baryons are still not accounted for \citep[]{Shull12} by observations \citep[however see][]{deGraaff2019, Tanimura2019, Macquart2020}. Moreover, the degree to which galaxy formation feedback impacts the low-$z$ IGM remains unclear \citep[][]{Khaire2023, Tilman23}, and simulations, even with extreme feedback, are still unable to reproduce the line-width distribution of the low-$z$ \lya forest \citep[][]{Gurvich17, Bolton_Gaikwad22, Khaire2023}. Given these challenges, studying the low-$z$ \lya forest becomes particularly interesting and is crucial in order to understand these discrepancies. The present work therefore focuses mostly on the low-$z$ \lya forest.  
 
Despite its potential, effectively extracting information from the \lya forest has proven to be challenging, especially when it is done via fitting Voigt profiles to the swath of absorption lines. Overlapping lines, varying signal-to-noise ratios, instrumental line-spread functions, and other systematic uncertainties can lead to parameter degeneracies, making it difficult to extract accurate physical information from the data. Usually, for dealing with a swath of \lya~lines, semi- or fully-automated codes are used to fit Voigt profiles, such as ``VoIgt profile Parameter Estimation Routine'' \citep[VIPER][hereafter, \citetalias{Prakash2017}]{Gaikwad2017}, BayesVP \citep{liang2017bayesvp}, GVPFIT \citep{Bainbridge2017}, and VoigtFit \citep{krogager2018voigtfit}; however, these are still computationally expensive for large samples. 

Moreover, efficient and automated analysis techniques are required to handle the increasing volume of data from modern surveys, both ongoing and upcoming, such as the Sloan Digital Sky Survey \citep[SDSS][]{SDSSpipeline2012}, Dark Energy Spectroscopic Instrument \citep[DESI, ][]{DESI2014}, William Herschel Telescope Enhanced Area Velocity Explorer \citep[WEAVE, ][]{WEAVE2016}, and 4-metre Multi-Object Spectroscopic Telescope \citep[4MOST, ][]{4MOST2012}, which are providing an unprecedented wealth of absorption spectra for analysis. In this regard, machine learning (ML) techniques offer a promising solution. 

ML algorithms, particularly those under the umbrella of deep learning \citep[][]{LeCun2015, goodfellow2016}, have demonstrated remarkable competency in pattern recognition, noise handling, and parameter estimation, making them well-suited to the challenges posed by the \lya forest.  ML models can learn the relationships between input data and desired outputs by training on large datasets of simulated or observed \lya forest. Furthermore, ML techniques can improve the efficiency of the fitting process. Automating the analysis using ML models reduces human intervention and subjective bias while enabling the rapid analysis of large datasets. 

In recent years, various applications of ML techniques have been developed to deal with a multitude of cosmological problems \citep[][]{Akhazhanov2021, Lee2021, Vattis2021, Liu2021, Olvera2021}. Deep learning or convolutional neural networks (CNNs) have been shown to be particularly powerful tools for cosmological data analysis. For example, \citet[][]{Parks2018} used a CNN to predict the \hi~column density of the damped \lya. \citet[][]{Huang2021} used a neural network to estimate the \lya optical depth values from noisy and saturated transmitted flux data in quasar spectra. \citet[][]{Veiga2021} used a deep neural network to infer the matter density power spectrum from the quasar spectra. \citet[][]{Cheng2022} also used CNN to identify the column density and Doppler widths of the \lya lines at high redshifts. Recently, \citet[][]{Stemock2023} also used deep learning to identify these parameters for Mg~{\sc ii} doublet absorption lines.

Motivated by these studies and with the aim of combining the power of ML and the rich information contained within \lya forest spectra, we developed FLAME (Fitting \lya Absorption lines using machine learning), which is the combination of a two-part algorithm that identifies the number of components in each \lya~absorption system and a fitting of the Voigt profiles to each of them. To train these ML models, we generated multiple simulated absorption lines with properties similar to low-$z$ \lya absorption lines observed with the Cosmic Origin Spectrograph (COS) on board the Hubble Space Telescope (HST). In addition to the reasons mentioned above, we focus our models exclusively on low-$z$ data to avoid complexities, because the \lya forest is less dense and shows minimal blending compared to high-redshift \lya forest regions. This allows easier isolation and fitting of each absorption line system. Nonetheless, even within low-$z$ data, both single and multiple components are present. Therefore, our two-part algorithm FLAME first determines the number of components present and then fits these identified components accordingly. 

To assess the robustness of our networks, we evaluated their performance on simulated, real observed \citep{Danforth2016}, and mock datasets. For the real observed dataset, we compared the model parameters with those derived using {\sc vpfit\footnote{\url{http://www.ast.cam.ac.uk/~rfc/vpfit.html}}} \citep{Carswell14} and VIPER. Our findings reveal that the neural networks demonstrate comparable performance while requiring significantly fewer computational resources.

The paper is organized as follows. We explain the terminology related to the ML algorithms in Sect.~\ref{sect-nn}. In Sect.~\ref{sect-data} we discuss the creation and preprocessing of the simulated data. We present the model and performances of the two ML algorithms in Sect.~\ref{sect-class} and Sect.~\ref{sect-regr}. 
In Sect.~\ref{sect-real}, we compare the accuracy of our ML algorithm to that of traditional algorithms on the observed data. In Sect.~\ref{sect-disscuss}, we discuss our key findings when using ML for Voigt profile fitting, and summarize our conclusions in Sect.~\ref{sect-summary}.

\section{Machine learning}
\label{sect-nn}
Machine learning is a branch of artificial intelligence that focuses on developing algorithms that learn from data and then make predictions without explicit programming. ML involves designing statistical and mathematical frameworks to uncover patterns, correlations, and trends within datasets autonomously. 

ML can be broadly classified as supervised or unsupervised. Unsupervised ML involves discovering patterns and structures in data without predefined labels. One of its applications is grouping similar data points based on certain features or characteristics. However, the ML model that finds a relation between the features in the measurements (training data) and its defining variables (labels) is known as the supervised ML. This trained model then predicts the label for any given set of measurements; the accuracy of this can be measured using validation data. In this paper, we only discuss supervised ML.

After defining the problem at hand, supervised ML can be summarized in six steps:
\begin{itemize}
    \item Preparing the training data that includes collecting and pre-processing the data (Sect.~\ref{sect-data}) with their corresponding labels. 
    \item Splitting the data into training and test/validation datasets. It is important that this splitting is random and that the training and testing cover a similar range of parameters to avoid possible extrapolation problems. 
    \item Generating an ML model and training it on the training dataset and predicting the labels for the testing dataset (Sect.~\ref{sect-class_model} and Sect.~\ref{sect-regr_model}). 
    \item Assessing the model's performance using the test or validation data by comparing the true and predicted labels.
    \item If the model's performance is unsatisfactory, adjusting hyperparameters, using different feature sets, or modifying the model architecture,
    \item Selecting the best-performing model, evaluating its performance (see Sect.~\ref{sect-class_result} and Sect.~\ref{sect-regr_result}).
\end{itemize}
These supervised ML algorithms commonly employ neural networks used across multiple applications.

\subsection{Neural networks and activation functions}

Neural networks work by sending information through layers of connected nodes or neurons, each contributing to the ability of the network to learn and make predictions. In these networks, every neuron is connected to all neurons in the next layer, and these connections have weights. These weights help determine how important each input is. At each $j^{th}$ neuron, the inputs are combined into a weighted sum, 
\begin{equation}
z_j = \sum_{i=1}^{n} (w_{ij} \cdot x_i) + b_j 
,\end{equation}
where $x_i$ is the input to the neuron from the previous layer, $w_{ij}$ as the weight of the connection between the $i^{th}$ neuron in the previous layer and the $j^{th}$ neuron in the current layer, $b_j$ as the bias term for the  $j^{th}$ neuron in the current layer, $z_j$ as the weighted sum of inputs to the $j^{th}$ neuron in the current layer, and $n$ is the number of neurons in the previous layer. This output is then processed by the activation function $f(z)$, which leads the output of the $j^{th}$ neuron in the current layer to be:  $a_j=f(z_j)$. This function $f(z)$ allows each neuron to introduce nonlinear transformations to its input, enabling the network to capture complex patterns in the data.  Without activation functions, the network would be limited to performing linear operations, resulting in a model incapable of capturing complex patterns in the data. Activation functions let the network handle a wide range of tasks and data patterns, making it capable of doing everything from simple classification to solving intricate problems across different areas. The combination of these weights, inputs, and activation functions is what enables neural networks to learn from various data and perform a broad spectrum of tasks, as discussed in \citet{Parhi2019}. 

We used two activation functions in this study: (a) Leaky Rectified Linear Unit (Leaky ReLU): $f(z)$ = max(0.01$z$, $z$), which returns $z$ if it receives any positive weighted sum input, but returns a small value of $0.01z$  if it receives any negative value of $z$. Therefore, it gives a positive output for negative values as well; and (b) Sigmoid: $f(z)$ = $1/(1+e^{-z})$. The value of this function exists between 0 to 1. This activation function is beneficial for models that predict the output as a probability, as the probability of anything exists between 0 and 1. We use sigmoid at the final layer of the classification model (Sect.~\ref{sect-class}). 

The hierarchical arrangement of layers enables the neural network to learn increasingly ``abstract representations'' as information progresses through the network. A network containing multiple fully connected layers is known as a "deep" neural network. However, to study the structure and patterns present in complex data, CNNs are designed. In the following subsection, we describe the CNNs.

\subsection{Convolutional neural network}
CNNs are specialized neural networks designed for processing data arranged in a grid-like structure, such as images or time series. They are particularly effective at detecting spatial relationships within the input data through a sequence of interconnected layers. In this study, we used CNNs to identify the number of blended \lya absorption lines and fit Voigt profiles to them.

At the heart of CNNs lie ``convolutional layers'', which utilize ``filters'' to apply convolutions to the input. A filter consists of multiple ``kernels'', with each kernel dedicated to a specific channel of the input. As these filters move across the input, they perform element-wise multiplications and summations, generating feature maps in the process. These feature maps are crucial for extracting various data characteristics, such as edges, textures, and patterns. A key hyper-parameter, ``stride'', governs the step size of the filter as it scans across the input. Padding, another important hyper-parameter, ensures comprehensive coverage of the input's edges, preserving the size of the input through the convolution process. This study employs the `{\sc same}' padding technique, ensuring that the dimensions of the output post-convolution remain consistent with the input dimensions. 

Following the convolutional layers, ``Pooling layers'' are used to reduce the spatial dimensions of the feature maps, thereby simplifying the information while retaining the most relevant features \citep{gholamalinezhad2020}. In this study, we use max pooling \citep[][]{Matoba2022}, a technique that identifies and retains the maximum value within a specific region defined by the kernel's coverage. This approach effectively captures the most prominent features within each region, enhancing the network's ability to understand spatial hierarchies and relationships.

In the following subsection, we explore how neural networks are trained and discuss the importance of loss functions in improving their accuracy.

\subsection{Neural network training and optimization}
After designing the neural network, we optimize the network's hyper-parameters by training the neural network to enable accurate predictions. The training procedure involves two main steps: forward propagation and backpropagation. 

During forward propagation, input ``training data'' flows through the network, with each neuron applying an activation function to the received signals, producing outputs. The computed outcomes are then compared to the true labels using a loss function, such as mean squared error (MSE) or cross-entropy, quantifying the error in predictions. For example, in this study, one of the loss functions used is binary cross-entropy (BCE), which has a functional form of:
\begin{equation}
H_p(q) = -\frac{1}{N}\sum_{i=1}^N y_i log(p[y_i]) + (1-y_i) log(1-p[y_i]),
    \label{eq-binary_cross}
\end{equation}
$H_p(q)$ signifies the entropy between the predicted probability distribution $p$ and the true distribution $q$. $N$ represents the dataset's total number of samples. $y_i$ is the ``true label'' of the i$^{th}$ sample in the dataset. It can be either 0 or 1 in a binary classification scenario. $p(y_i)$ is the predicted probability that the i$^{th}$ sample belongs to class 1 (or has label 1). The other loss function used in this study is the mean squared error (MSE),
\begin{equation}
MSE = -\frac{1}{N}\sum_{i=1}^N (y_i-\hat{y_i})^2
\label{eq:mse}
,\end{equation}
where $y_i$ and $\hat{y_i}$ represent the true and predicted labels for the $i^{th}$ sample, respectively.

Back-propagation involves propagating the error backward through the network to compute gradients of the loss function with respect to the network's hyper-parameters. These gradients provide valuable information on adjusting the parameters to minimize the error. Popular optimization algorithms, like Adagrad, RMSprop, Stochastic Gradient Descent (SGD), and ``Adam'' \citep[][]{kingmaandba}, update the parameters iteratively to minimize these gradients. One of the crucial hyperparameters in ML algorithms is the learning rate. It determines the step size at which the model parameters are updated during the optimization process.
This study uses the Adam optimizer, which is a more efficient alternative to the other methods, to adjust the model weights. 

During training, the network updates its parameters by utilizing smaller subsets or batches of the dataset. This study uses smaller ``batch sizes'' that allow more frequent updates to the network's parameters, leading to faster convergence and mitigating memory limitations \citep[][]{You2017}. Another hyperparameter is the number of ``epochs determining how often the network will iterate over the training dataset. We define the epochs for this study by implementing an ``early stopping technique''. This approach halts the training process when the validation performance no longer improves, helps prevent overfitting, and conserves computational resources. 

Neural network training relies on a labeled dataset, careful selection of architecture, regularization techniques to prevent overfitting, and hyperparameter tuning to achieve optimal results. We iterate the hyperparameters mentioned above after carefully selecting the data and model. Then, we choose the hyperparameters that produce satisfactory results, measured using an evaluation metric. Below, we discuss the evaluation metrics used in this study. 

\subsection{Evaluation metrics}
After training the network and finding the best hyperparameters, we validate the model using a test or validation dataset. The evaluation metrics of this test dataset ensure the model is unbiased and test its performance. The labels of the testing dataset are called the ``true'' labels, and the predictions made by the model are known as the ``predicted'' labels. In this study, we use two models as described in Sect.~\ref{sect-class_model} and ~\ref{sect-regr_model};
(i) Binary classification algorithm - to classify the number of absorptions into single and double lines, and (ii) Regression algorithm - to identify the Voigt profile parameters of the absorption lines. Therefore, we require two evaluation metrics.

In the binary classification algorithm, there are two classes: positive and negative. To test the performance of the binary classification algorithm \citep{Hossin2015}, we calculate the accuracy,
\begin{equation}
    accuracy = \frac{TP+TN}{TN+ FP+ FN+ TP}, 
    \label{eq-accuracy}
\end{equation}
where TP is true positive, TN is true negative, FP is false positive, FN is false negative. The true positives and negatives imply that the model's outcome correctly predicts the positive and negative classes. The false positive means the model's outcome incorrectly predicts a positive class for an actual negative class. The false negative means the model's outcome incorrectly predicts a negative class for an actual positive class.

Other parameters to identify the robustness of the classification algorithm are as follows:
\begin{itemize}
\item{\bf Sensitivity}, also known as recall, assesses a model's ability to correctly identify positive instances, focusing on minimizing false negatives.
\item{\bf Specificity}, also known as true negative rate, gauges a model's capacity to recognize negative instances, aiming to minimize false positives accurately.
\item{\bf Precision} quantifies the proportion of correctly predicted positive cases among all instances predicted as positive, emphasizing the minimization of false positives.
\item{\bf Negative predictive value} evaluates the proportion of accurately predicted negative cases among all instances predicted as negative, giving insight into the model's ability to avoid false negatives.
\end{itemize}
The precision and sensitivity lead to the F1-Score. The F1-score is the harmonic mean of precision and recall as given by\begin{equation}
    \text{F1-Score} = 2 \times \frac{Precision \times Recall}{Precision + Recall}.
    \label{eq-f1}
\end{equation}

The F1-score is a single metric that balances both precision and recall and is especially useful when the class distribution is imbalanced. The F1-score ranges between 0 and 1, with higher values indicating better performance. An F1-score of 1 indicates perfect precision and recall. In practical terms, this indicates that the test has achieved the highest possible accuracy, with no false positives or false negatives. However, an F1-score of zero arises when either precision or recall is zero, indicating that the test's accuracy is at its lowest. During the regression analysis (as described in Section.~\ref{sect-regr}), we predict values from the model and compare them to the true labels of the dataset. To test the accuracy of the regression analysis, we use the MSE. We also calculate the 90 and 68 percentile values of the absolute differences in the true and predicted values to understand the data distribution and its concentration. We use percentiles because they are less sensitive to extreme values or outliers since they're based on rank order rather than actual values. 

\begin{figure}[!t]
    \centering
    \includegraphics[width=5.5cm,height=8cm,trim={6cm 0cm 6cm 0cm}]{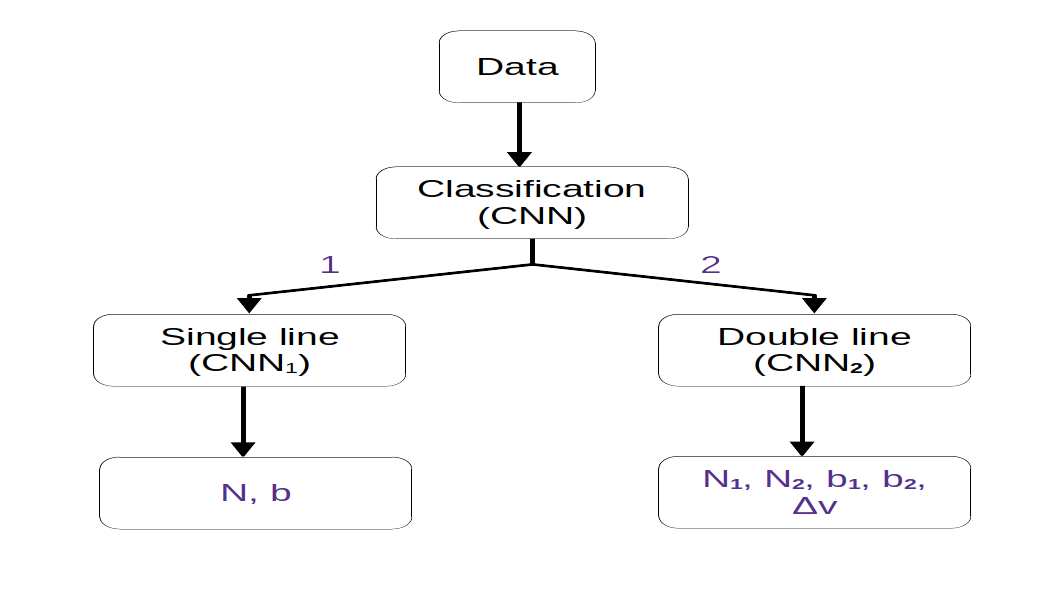}
     \caption{The flowchart outlines the sequence of three neural networks comprising FLAME.  The input training data to the classification algorithm is the normalized flux with 301 array size labeled as 1 or 2 absorption lines. The same input data is then fed to different regression algorithms based on the number of absorption lines. For this algorithm, the labels with the input dataset are the physical parameters like column density and Doppler width. }
    \label{fig-flowchart}
\end{figure}

Figure~\ref{fig-flowchart} shows the flowchart sequence of the two models used in this study. The first algorithm is a CNN that classifies the number of absorption lines into either single or double. The choice of only two states is dictated by the low-$z$ dataset we used (see Section.~\ref{sect-data}). Once the classification is performed, we use the second algorithm, which consists of two CNNs, to estimate the features/parameters, one for single lines and the other for double lines. These networks are created using the TensorFlow\footnote{\href{https://www.tensorflow.org/api_docs/python/tf/keras}{https://www.tensorflow.org/api\_docs/python/tf/keras}} interface \citep[][]{tensorflow2015}. We construct a simulated dataset of low-$z$ \lya lines and use it to train the networks created here. We outline the details of the dataset in the next section.

\section{Training and testing dataset}
\label{sect-data}
The training dataset is the initial input to the network during the training process and plays a crucial role in shaping the network's ability to learn and make predictions. Therefore, we require a well-constructed training dataset that covers a range of parameter space and represents the real-observed data. However, due to the limited sample available of the low-$z$ \lya absorption lines, we generate our training dataset by simulating Voigt profiles. In order to model realistic simulated \lya lines, we use the general properties of the observed low-$z$ \lya data. Below, we explain the properties of the low-$z$ data and how it was used in simulating the lines for training.

\subsection{Observational data description}
\label{sect-obs_data}
We aim to apply ML techniques to fit low-redshift \lya lines, drawing on the extensive survey of the low-redshift IGM at $z < 0.48$ conducted by \citet{Danforth2016}, henceforth referred to as \citetalias{Danforth2016}. This survey uses 82 high signal-to-noise ratio (S/N) quasar spectra observed with the HST/COS in the far-ultraviolet (FUV) band using medium-resolution gratings G130M and G160M ($R \sim 18000$, $\Delta v \sim 18$ km s$^{-1}$) across different lifetime positions to cover a wavelength range of 1030 \AA \ to 1800 \AA. \citetalias{Danforth2016} combined data from both gratings whenever available, applied manual continuum fitting to each spectrum, and cataloged all absorption lines, including those intervening, associated, and arising from the Milky Way's interstellar medium.

\citetalias{Danforth2016} determined locations of absorption lines using a crude significance level vector SL($\lambda$) $= W(\lambda) \bar{\sigma}(\lambda) > 3$, where $W(\lambda)$ represents the equivalent width vector and $\bar{\sigma}(\lambda)$ denotes the error vector in regions without lines. Following this localization, a standard procedure was employed to identify lines by using coincident higher-order lines or lines from different ions. They identified a total of 2611 intervening \lya\ lines and modeled them with Voigt profiles. The line list tables from \citetalias{Danforth2016}, publicly available in the high-level science product at the Mikulski Archive for Space Telescopes\footnote{\url{https://archive.stsci.edu/prepds/igm/}}, include details of these fits, such as the Doppler width ($b$), redshift ($z_{abs}$), and neutral hydrogen column density (N$_{\textsc{HI}}$), along with their associated errors.

For the present study, we selected 1917 \lya lines that are not blended with metal lines or higher-order lines from the \citetalias{Danforth2016} catalog. According to \citetalias{Danforth2016}'s fits, 81.2\% (1557) of these lines are single lines, 14.9\% (286) are doublet structures, and the remaining 3.8 \% (74) consist of three or more lines. We used the parameters of these lines to generate the simulated training dataset described in the following section.

\begin{figure*}[!t]
    \centering
    \includegraphics[width=8cm,height=7cm]{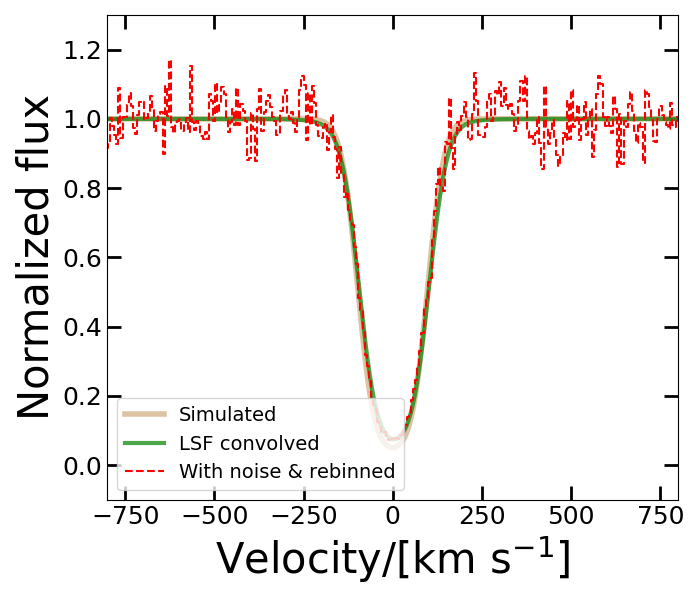}
     \includegraphics[width=9cm,height=7cm]{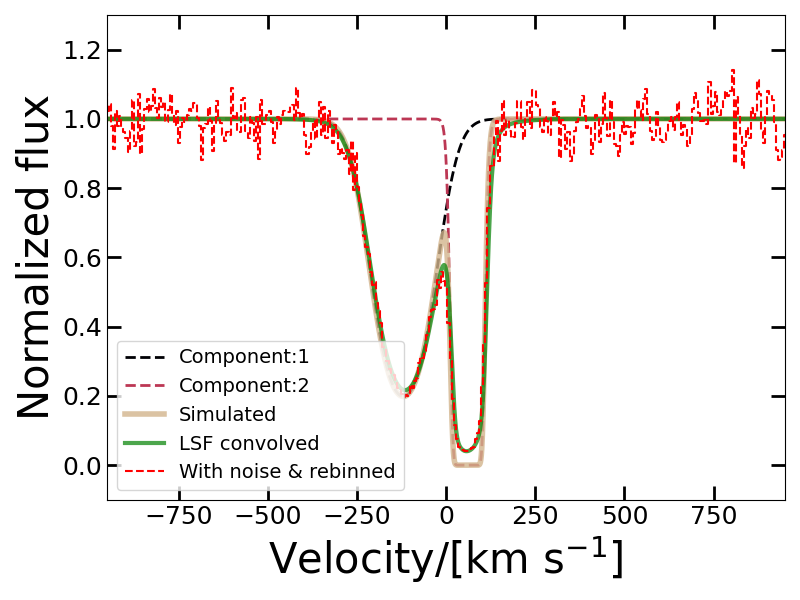}
     \caption{Simulated Voigt profiles for single and double absorption lines in brown color. The green lines show the absorption lines after convolving with the HST's tabulated LSF. The red dashed line shows the absorption line after adding the Gaussian noise and rebinning it to a similar velocity frame to the HST data. This red dashed line represents the typical absorption line used as the training dataset in this study. {\it Left panel:} Brown line shows the simulated Voigt profiles for a single absorption line with N$_{\rm HI}$ = 10$^{14}$ cm$^{-2}$, b= 80 km s$^{-1}$ and S/N=25.  {\it Right panel:} Same as the left panel but a simulated double absorption line. The dashed lines show two simulated single absorption lines (N$_{\rm HI}$ = 10$^{14.29}$ and 10$^{15.06}$ cm$^{-2}$, b= 92.67 and 25.43 km s$^{-1}$) that are shifted by $\pm \Delta v/2 \sim 75$ \kms value and the combined profile is shown in brown color. The training dataset for double absorption lines is shown in red dashed lines.} 
    \label{fig-spectra}
\end{figure*}

\begin{figure*}
    \centering
    \includegraphics[width=0.97\textwidth]{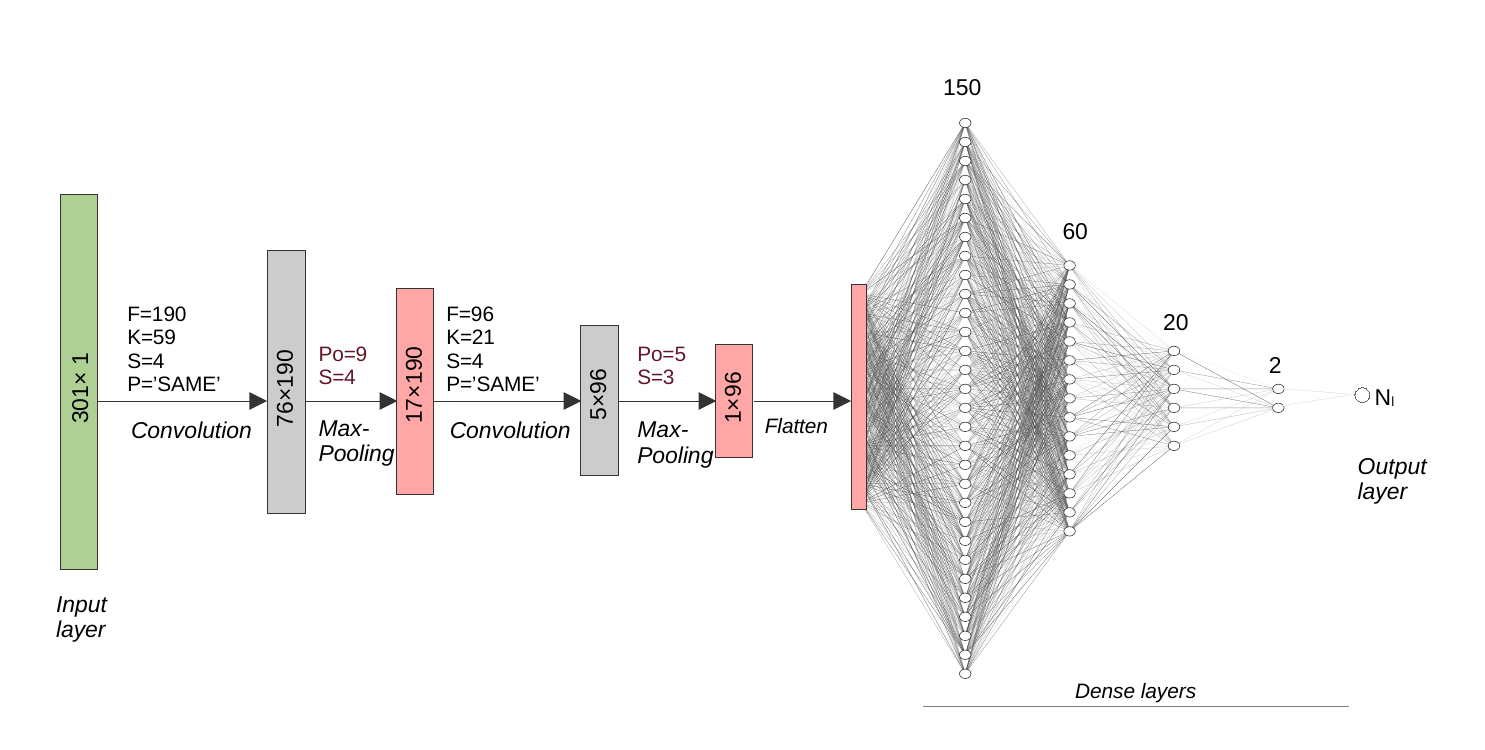} 
    \caption{Schematic diagram of the CNN architecture to classify the number of absorption lines. The notation used here is F - filter size, K - kernel size, S - strides, P - Padding, and Po - Pooling size. The number of neurons in the dense layers is written at the top. The last layer has one output varying between 0 and 1. The values $<0.5$ are assigned N$_l=1$, and values $\geq0.5$ are assigned N$_l=2$.}
    \label{fig-classify_network}
\end{figure*}

\subsection{Simulated data}
\label{sect-simudata}
The simulated \lya absorption lines are generated with randomly selected (from a uniform distribution) column density (N$_{\rm HI}$/[cm$^{-2}$]) and Doppler width ($b$/[km s$^{-1}$]) and combined it with instrumental properties of HST/COS data. The range for each parameter includes the minimum and maximum values (for 98.5\% data avoiding the outliers in each parameter) of the HST data, as mentioned above.  Since the real observed dataset (Sect.~\ref{sect-obs_data}) has $<4\%$ lines with $>2$ components, our simulated dataset is limited to single and double lines. The parameter ranges are as follows:

\begin{itemize}
  \item Doppler width: $b$ (\kms) = 5-100,
  \item Column density: log  N$_{\rm HI}$ ({cm}$^{-2}$) = 12-17,
  \item Signal-to-noise ratio: S/N = 5-100,
  \item Central wavelength: C$_\lambda$ (\AA) = 1220-1800.
\end{itemize}

Given these parameters space, we used the following steps to create the dataset for single Voigt profiles:
\begin{enumerate}
\item Generate the Voigt profiles using the `Faddeeva function', for a randomly selected value of $b$ and N$_{\rm HI}$ from the above range, resulting in optical depth ($\tau$) over a velocity resolution of 0.6 \kms. 
\item Convolve the simulated flux (F= $e^{-\tau}$) with the tabulated line spread function (LSF) of the HST COS spectrograph.\footnote{\href{https://spacetelescope.github.io/COS-Notebooks/LSF.html}{https://spacetelescope.github.io/COS-Notebooks/LSF.html}} The Line Spread Function (LSF) varies with each grating and is wavelength-dependent. Thus, the convolution depends on both the observed wavelength ($z_{abs}$) and the grating used.
\item We randomly select a central wavelength value (C$_\lambda$) from above range. For C$_\lambda$ $> 1450\AA$, a value of $z_{abs}$ or $z$ is selected randomly from a uniform sample between 0.2 to 0.47, and the grating is G160M. However, for central wavelength $\leq 1450\AA$, $z_{abs}$ are randomly selected from a uniform sample between 0.005 to 0.2, and grating is G130M. The convolution is performed in the observed frame.
\item Resample the convolved data to a similar wavelength scale ($\Delta_\lambda$ = 0.0299 \AA\ i.e.,$\Delta v =  6$ \kms) with which \citetalias{Danforth2016} resampled their combined spectra.
\item Add Gaussian random noise to the simulated Voigt profile with a signal-to-noise ratio (S/N) selected from 5-100.
\item Convert the rest-wavelength to the $\Delta v$.
We align the absorber at the center by selecting a chunk of 301 pixels and pad the rest of the chunk with continuum flux = 1.
\end{enumerate}
After creating a sample of single absorption lines, we used the following procedure to 
generate a sample representing double lines, which simulate blends of two components within an 
absorption system. Initially, we randomly selected two simulated Voigt profiles with optical depths $\tau_1$ and $\tau_2$ from the first step. 
Each absorption line was then shifted by $\pm \Delta v/2$, where $\Delta v$ is randomly chosen from 
the range of 5 to 350 km s$^{-1}$. This ensures that the center of the absorption line does not 
align closely with the spectral edges. Subsequently, the combined optical depth ($\tau = \tau_1 + 
\tau_2$) was converted into flux (F = $e^{-\tau}$), serving as the input for the second step. The 
procedures outlined above are similarly applied throughout.

Figure~\ref{fig-spectra} illustrates two examples of our simulated absorption lines. The left panel of Fig.~\ref{fig-spectra} shows a single Voigt profile for a given $b$, N$_{\rm HI}$ and $z_{abs}$; the right panel shows an overlapping double Voigt profile structure with velocity separation ($\Delta v$) for a given pair of $b$ and N$_{\rm HI}$. A solid brown line in Fig.~\ref{fig-spectra} shows the simulated line generated in step (i). After convolving the line with COS LSF corresponding to step (ii) is shown in the green line. The red dashed line shows the final absorption lines after rebinning and adding the Gaussian noise.

The above-generated dataset is divided into 80\% training and 20\% testing datasets, covering the same parameter ranges. The ML models use the training dataset as input and evaluate their accuracy on new, unseen data with the testing dataset.

\subsection{Mock data}
\label{sect-mock}
Ideally, the accuracy of the ML algorithm should be consistent with simulated and real testing datasets. In the case of any disagreement, generating a mock dataset can effectively resolve any issues with the algorithm's performance or the creation of the simulated testing dataset. Therefore, complementing our study of the ``real'' data, we also generated a set of ``mock'' testing data. These mock \lya lines mimic the real dataset generated by the procedure discussed in Sect.~\ref{sect-simudata}. The mock absorption lines have exactly the same physical parameters ($b$, log N$_{\rm HI}$, S/N, $z,$ and C$_\lambda$) as the real dataset. In subsequent sections, we also present an assessment of the ML algorithm's performance on this realistic mock data and demonstrate its reliability.

\begin{figure}
    \centering
    \includegraphics[width=0.4\textwidth,trim={0cm 0cm 0cm 0cm}]{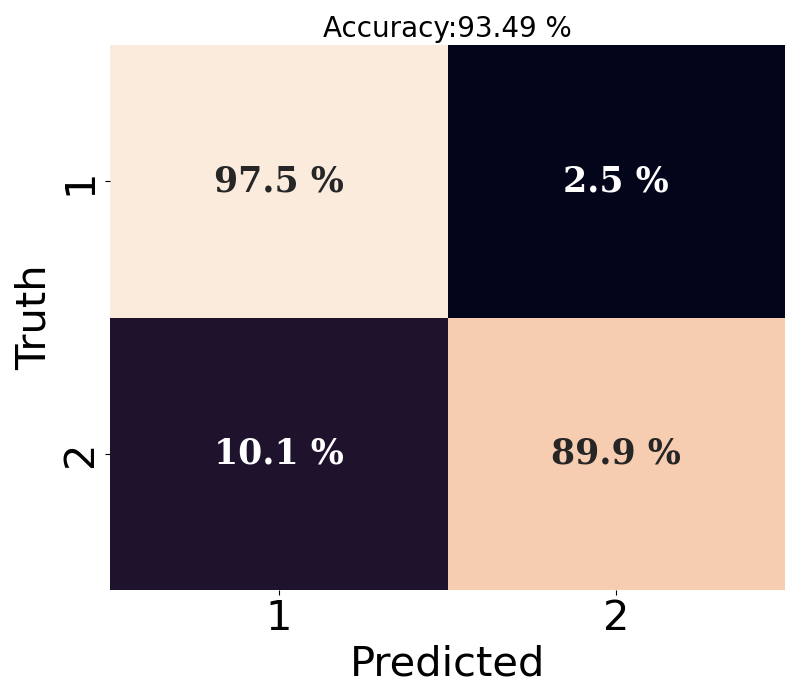}   
    \caption{Confusion matrix for the predictions for the number of absorption lines using the CNN (as shown in Fig.~\ref{fig-classify_network}). The CNN was trained on 1.6 million lines and tested on 400K samples, with an equal number of single and double lines. We find the Sensitivity=97.47\%, Specificity=89.92\%, Precision=89.64\% and Negative Predictive Value=97.55\%.} 
    \label{fig-confusion_matrix}
\end{figure}

\section{Classifying the number of absorption lines}
\label{sect-class}

\subsection{CNN architecture}
\label{sect-class_model}
This section introduces an ML algorithm based on CNN architecture that is specifically designed for binary classification to identify the number of absorption lines. The input data consists of a training dataset generated in Sect.~\ref{sect-data}, comprising 301 pixels representing simulated absorption lines. The model's output is a single value ranging between 0 and 1, where values $<$ a threshold value indicate a single absorption line ($N_l$=1) and $\geq$ threshold value corresponds to a double absorption line ($N_l$=2). We tested model outputs with various threshold values and found that $0.5$ gives an unbiased result.

The CNN architecture shown in Fig.~\ref{fig-classify_network} includes an input layer with 301 neurons, two convolutional layers, two max-pooling layers, and four fully connected dense layers with a decreasing number of neurons. The activation function employed throughout the model is the Leaky ReLU, except the sigmoid function in the last layer. We use the binary cross-entropy loss function {\bf (Eq.~\ref{eq-binary_cross})} and the Adam optimizer with a learning rate of $10^{-3}$ for optimization.

The CNN is trained on 1.6 million absorber samples and tested on 4$\times$ 10$^5$ samples (see Sect.~\ref{sect-data}), with an equal number of single and double absorption lines. We apply batch propagation with batch size 100 and implement early stopping criteria with a patience of 20 epochs. To ensure robustness and accurate identification of the number of Voigt profiles in a chunk of 301 pixels, we carefully fine-tune the parameters of the binary classification algorithm (Fig.~\ref{fig-classify_network}). Our objective is to achieve an accuracy of over 90\% (see Eq.~\ref{eq-accuracy}) and F1-score greater than 0.8 (see Eq.~\ref{eq-f1}). 

We also tested with other ML algorithms, like random forest classifiers \citep[][]{breiman2001} and support vector machines \citep[][]{cortes1995}. However, we found that the F1-score for all other classifiers was less than 0.8. 

\begin{figure}[!t]
    \centering
    \includegraphics[width=0.52\textwidth,trim={1cm 0cm 0cm 0cm}]{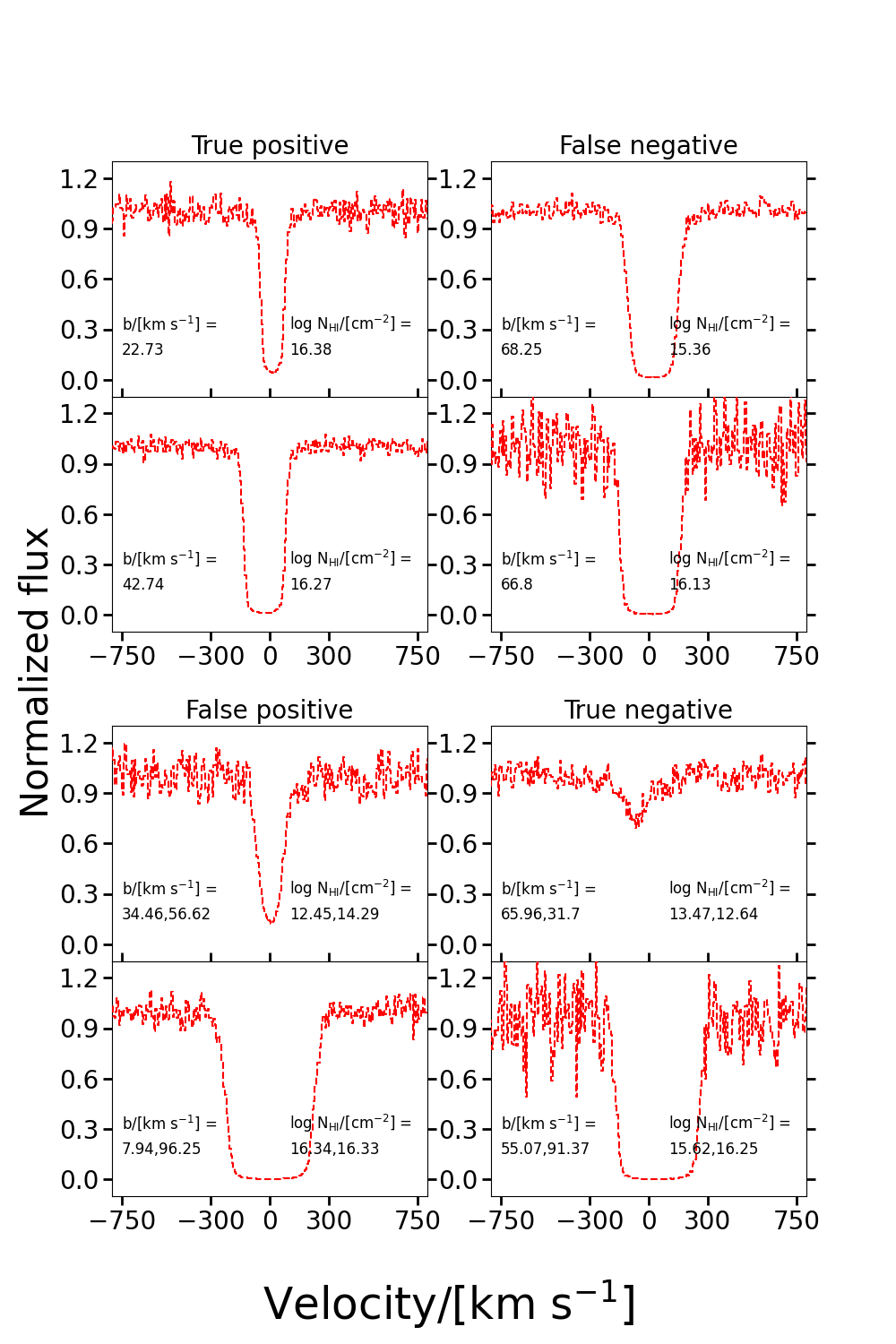}
    \caption{Each panel shows two examples from the four classes of Fig.~\ref{fig-confusion_matrix}. The top left and bottom right panels show the correctly predicted single and double absorption lines. The top right and bottom left show the misclassified examples.}
    \label{fig-classify_examples}
\end{figure}

\begin{figure*}
     \centering
     \includegraphics[width=0.32\textwidth]{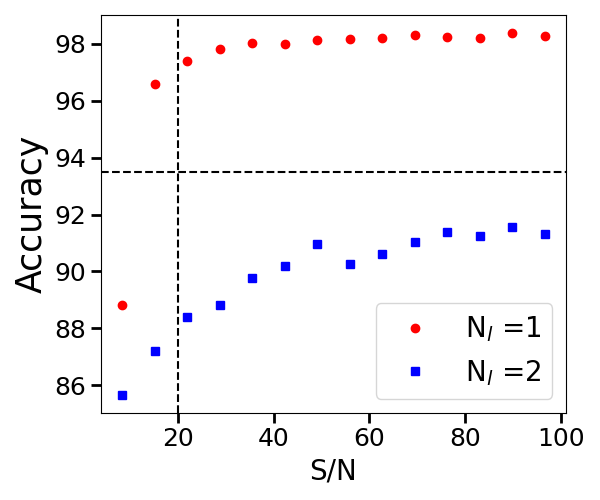}
     \includegraphics[width=0.32\textwidth]{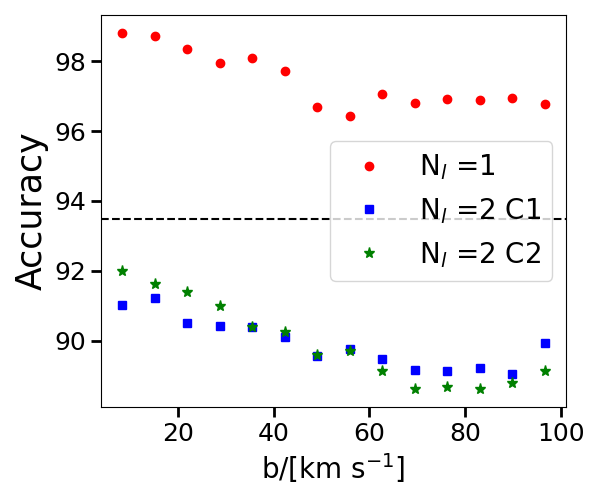}
     \includegraphics[width=0.32\textwidth]{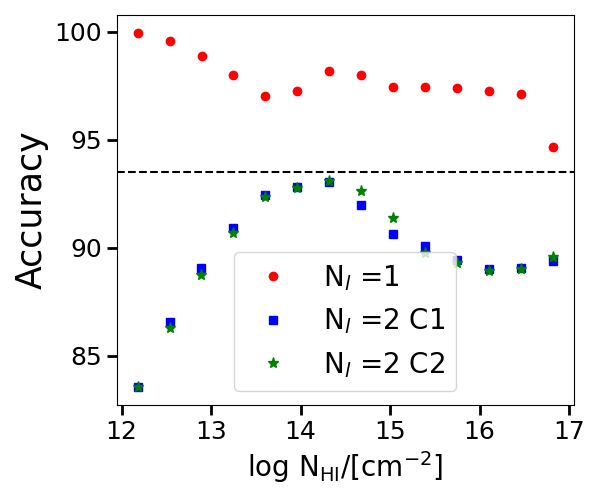}
      \caption{Comparison of classification accuracy versus signal-to-noise ratio, Doppler width ($b$), and column density (N$_{\rm HI}$) for simulated data. $N_l=1$ represents the single absorption lines, and $N_l=2$ C1 and $N_l=2$ C2 represent the first and second components of double absorption lines, respectively.}
     \label{accuray_snr}
\end{figure*}

\begin{figure*}
    \centering
    \includegraphics[width=0.60\textwidth,trim={6cm 0cm 6cm 0cm}]{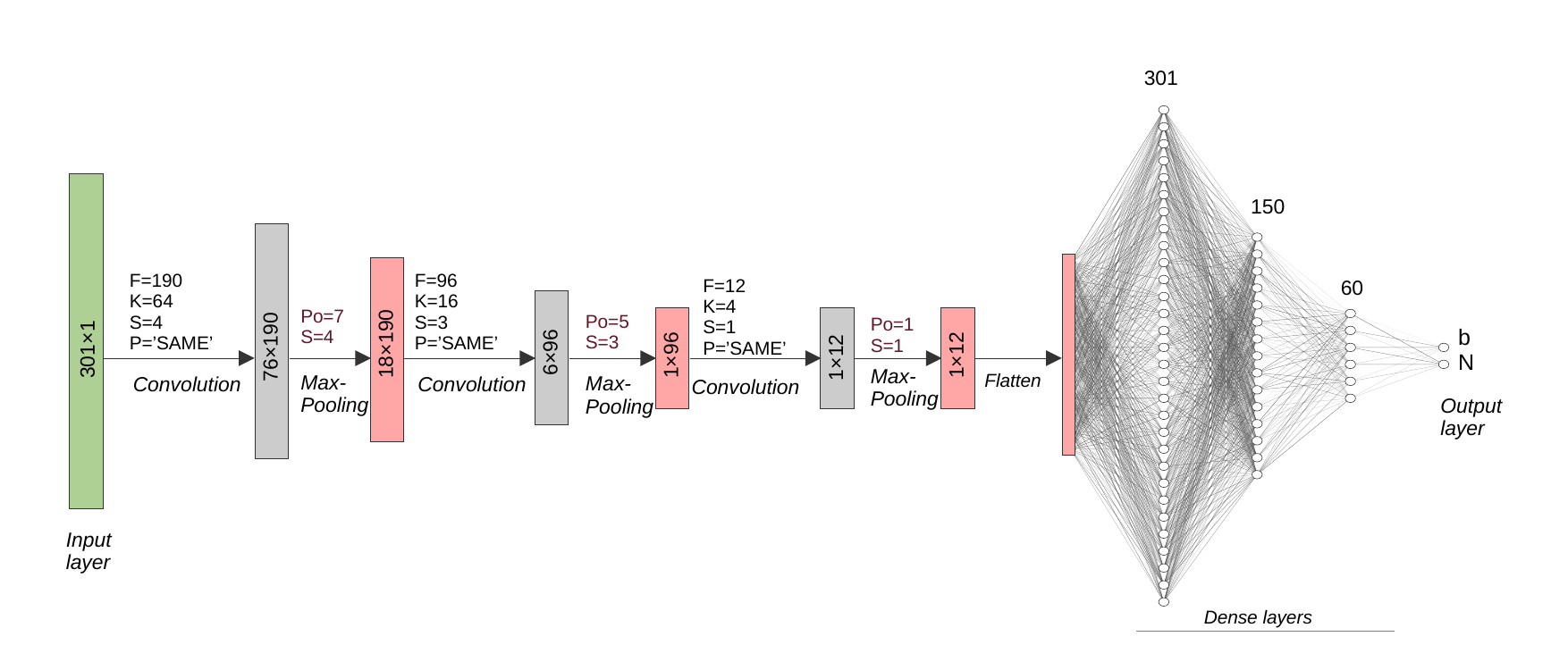}   
    \caption{Schematic diagram of the CNN model to predict the $b$ and N values for a single absorption line. The CNN comprises three one-dimensional convolutional layers, each followed by three max-pooling layers; after flattening, the output is input to three dense layers with a decreasing number of neurons and two neurons in the output layer.}
    \label{fig-parameter_network}
\end{figure*}

\subsection{Classification of the number of \lya absorption lines}
\label{sect-class_result}
Figure~\ref{fig-confusion_matrix} shows the performance of the binary classification algorithm computed for the simulated test dataset. In this study, the true positives and true negatives are ``true-single'' and ``true-double'' lines, respectively. The simulated test sample has an impressive TP rate of 97.5\%, accurately identifying $N_l$=1 absorption lines. The FP rate was 10.1\%, reflecting a moderate number of misclassifications of $N_l$=2 absorption lines as $N_l$=1. However, the TN rate was high at 89.9\%, indicating a highly reliable identification of $N_l$=2 absorption lines. We also noted a small FN rate of 2.5\%, indicating a low number of misclassifications of $N_l$=1 absorption lines as $N_l$=2. The normalized confusion matrix (Fig.~\ref{fig-confusion_matrix}) demonstrates the binary classification algorithm's ability to distinguish between the two absorption line categories. The caption mentions the values of sensitivity, specificity, precision, and negative predictive values. This leads to an F1-score (see Eq.~\ref{eq-f1}) of 0.93, suggesting the model accurately identifies single and double absorption lines. 

Figure~\ref{fig-classify_examples} shows two representative examples from each category. Even by visual inspection, it becomes evident that the misclassified absorption lines also appear visually ambiguous. We evaluated the model's performance across different parameters, including S/N,  $b$-parameter, and N$_{\rm HI}$ as shown in Fig.~\ref{accuray_snr}. As expected, accuracies are notably lower for small values of S/N.  This effect is also evident from the examples FP and FN in Fig.~\ref{fig-classify_examples}. However, excluding cases with S/N$<$20 in the lower percentile consistently yields accuracies above 98\% and 91\% for single and double lines, which is promising. We also find that the accuracy for single lines is consistently better than the accuracy for double lines. We also find that the accuracy decreases slightly with increasing Doppler width. For broad absorption features, it is reasonable to expect that the CNN may encounter greater difficulty in discerning whether it originates from a single line or double lines. This pattern is similar for column density for single lines but not for double lines. The accuracy for lower column densities is influenced by S/N, as absorbers with low column densities may be more susceptible to being hidden within noise. On the other hand, for higher column densities, the saturation of absorption lines can limit the accuracy of classification. Even if one of the lines in double lines is saturated, regardless of the column density of the other line, classification becomes challenging. 
The evaluation of the simulated test sample and the visual examination of classification examples highlight the CNN-based algorithm's effectiveness and reliability in absorption line classification tasks.

\begin{figure}
 \centering
 \includegraphics[width=0.23\textwidth]{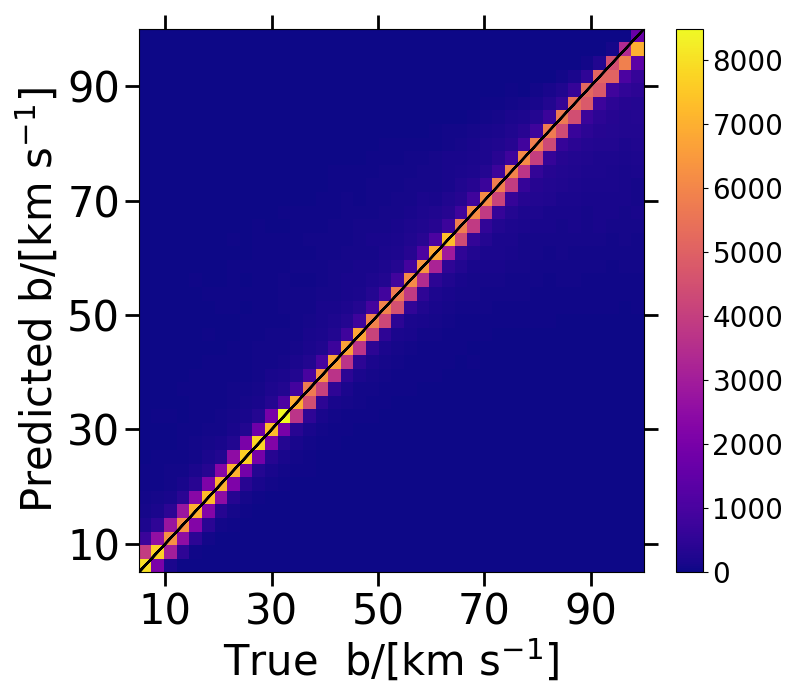}
 \includegraphics[width=0.23\textwidth]{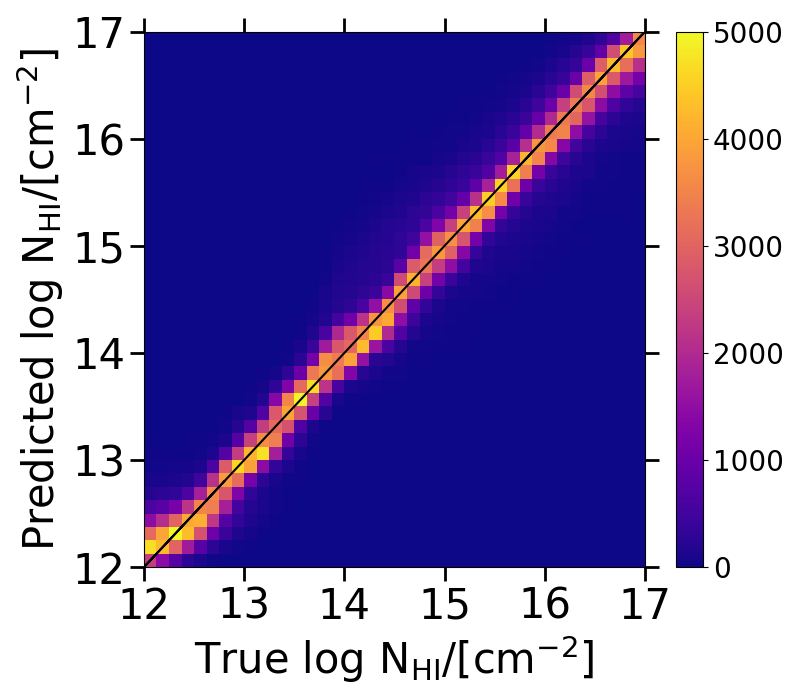}
 \includegraphics[width=0.23\textwidth]{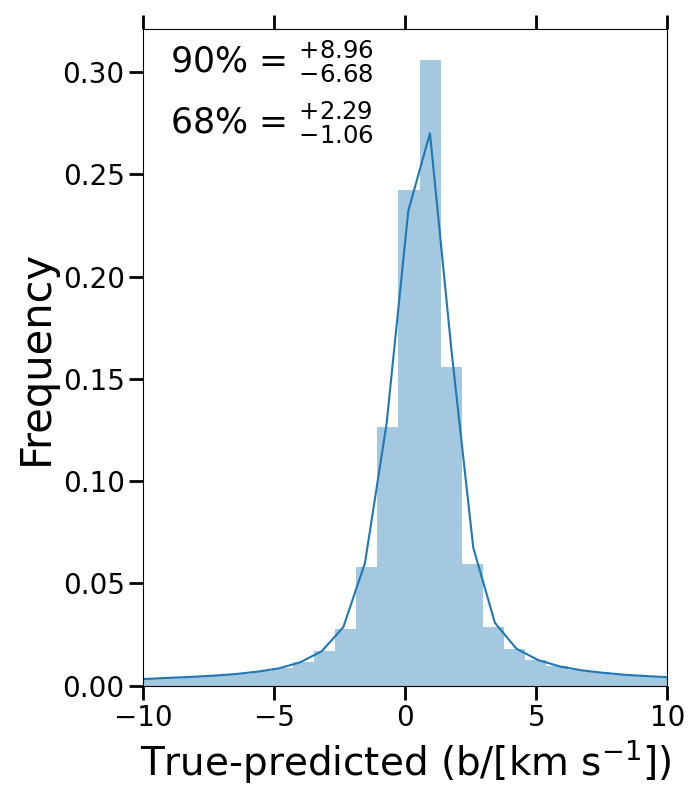}
 \includegraphics[width=0.23\textwidth]{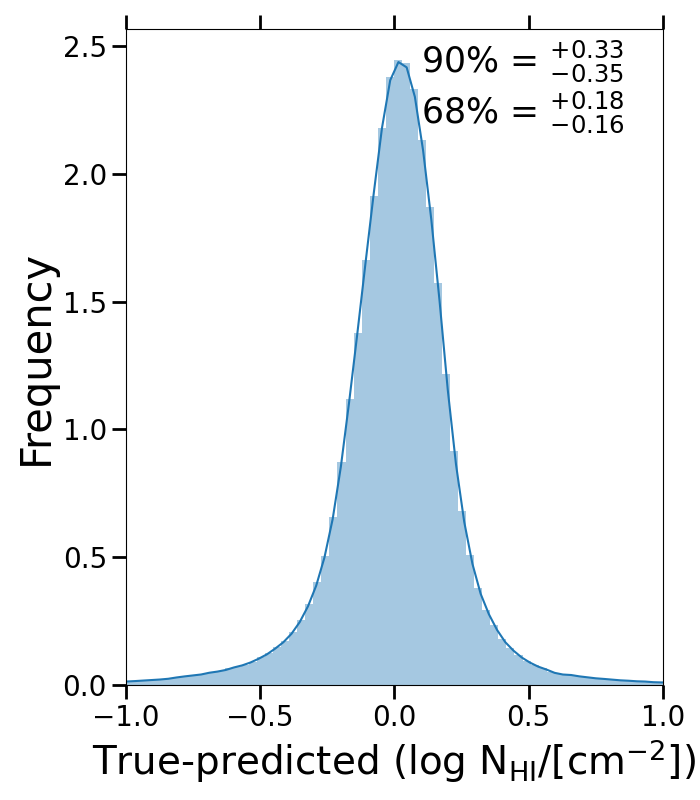}
\includegraphics[width=0.23\textwidth]{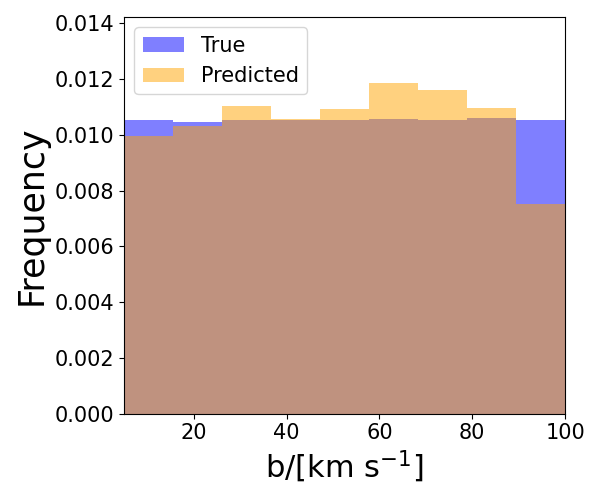}
\includegraphics[width=0.23\textwidth]{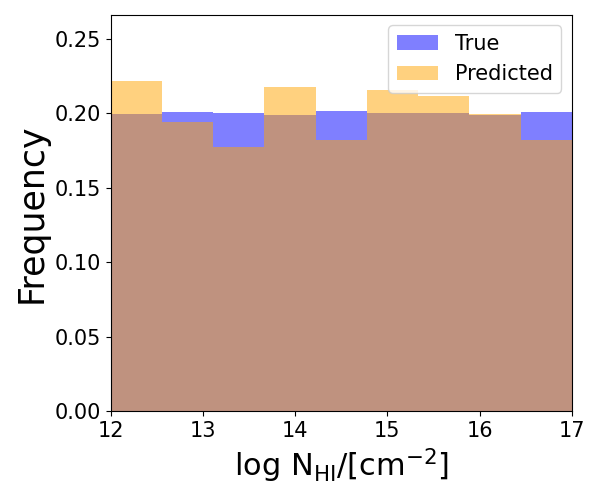}
  \caption{Comparison and evaluation of the predicted and true parameters for simulated test single absorption line. The upper panel compares the actual and predicted values for the two parameters, $b$ and N$_{\rm{HI}}$. The middle panel exhibits the normalized distribution of the differences between the predicted and true values, with markers for the 90\% and 68\% percentiles. The bottom panels show the normalized histogram of the CNN-predicted values in comparison to the true labels.}
    \label{fig-result2}
\end{figure}

\section{Regression analysis}
\label{sect-regr}
\subsection{Algorithm}
\label{sect-regr_model}
In this section, we outline the regression analysis ML algorithm designed to determine the physical properties of absorption lines.  We train this network on the simulated dataset generated in Sect.~\ref{sect-data}, similar to the previous model. The primary output predictions are the parameters characterizing the absorption lines. For single lines, these parameters encompass log N$_{\rm HI}$  (or log N) and $b$, while for double lines, they extend to log N$_1$, $b_1$, log N$_2$, $b_2$  (the subscripts represent components), and $\Delta v$. 

To build the regression analysis, we initially experimented with fully connected deep neural networks with varying numbers of hidden layers and other units to develop the ML algorithm. However, we found that a large input dataset was required for the algorithm to learn effectively, which was computationally expensive. Therefore, to optimize the model's learning efficiency, we use CNNs for parameter estimation. The architecture, illustrated in Fig.~\ref{fig-parameter_network}, comprises of three convolutional layers, each followed by a max pooling layer, with decreasing size of hyper-parameters, connected to three dense layers. The output layer results in the estimate of $b$ and N$_{\rm{HI}}$.

The single absorption lines dataset consists of 2.5 million samples randomly divided into 80\% as training and 20\% as testing datasets.  The data size was selected to minimize computational time and achieve higher accuracy. All the convolutional and dense layers are activated using the LeakyReLU activation function. We used the Adam optimizer with a $10^{-4}$ learning rate. The model undergoes training in batches of ten instances, employing early stopping with the patience of 100 epochs. Training halts after 128 epochs based on the specified early stopping criteria. The weights are updated based on the MSE loss function (Eq.~\ref{eq:mse}). The hyper-parameters were varied in order to select those hyper-parameters that resulted in the minimum discrepancy between the true and predicted values.

For double absorption lines, the architecture was similar to the above CNN (Fig.~\ref{fig-parameter_network}) with minor modifications:
\begin{itemize}
    \item To obtain similar outcomes, the input data consisted of a sample size of 4 million, split into two sets: 80\% for training and 20\% for testing. 
    \item The output layer has five values ($b_1$, $b_2$ , log N$_1$, log N$_2$ and $\Delta v$).
    \item  Using early stopping criteria, we trained the algorithm for 430 epochs.
\end{itemize}

 \begin{figure}
     \centering
     \includegraphics[width=0.5\textwidth]{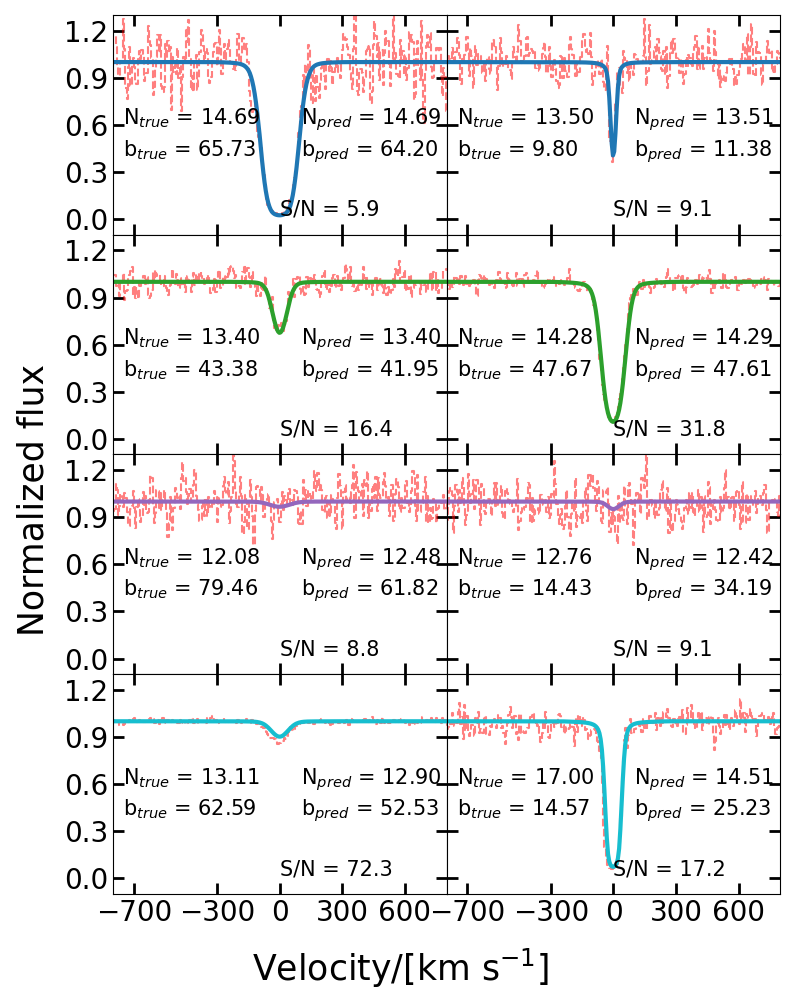}
      \caption{Examples of the simulated test single absorption lines and the corresponding  Voigt profile predicted by the CNN model (in colors). The input data, consisting of 301 pixels generated in Sect.~\ref{sect-data}, are shown as red dashed lines. The true and predicted parameters are written in each panel. The figures in the upper panel demonstrate precise physical parameter predictions made by the CNN for low-S/N data (blue), while the second panel displays higher-S/N data (green). The two lower panels illustrate two instances of an inaccurate prediction made by CNN for low- (purple) and high-S/N (cyan). To be considered an accurate prediction, the criterion is |N$_{true}$-N$_{pred}| < $ 0.33 cm$^{-2}$ ($\sigma_{90N}$ in log scale)  and |b$_{true}$-b$_{pred}| < 7.8 $ \kms ($\sigma_{90b}$). Alternatively, |N$_{true}$-N$_{pred}| \geq 0.33$ cm$^{-2}$ and |b$_{true}$-b$_{pred}| \geq 7.8$ \kms serve as the criterion for inaccurate prediction. }
     \label{fig-cnn_n1_example}
 \end{figure}
\begin{figure*}
    \centering
    \includegraphics[width=6cm,height=5cm]{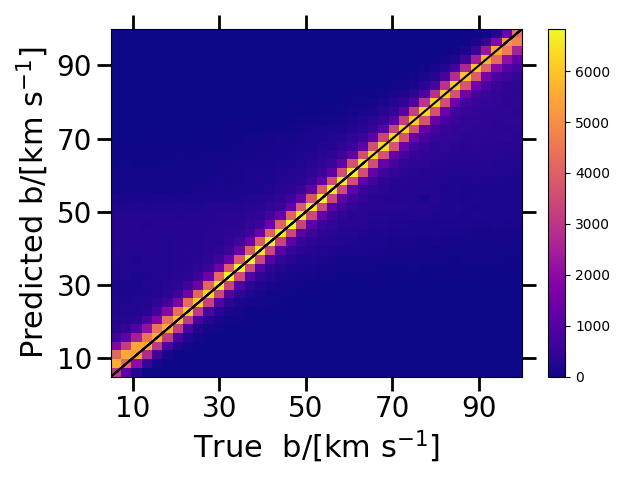}  \includegraphics[width=6cm,height=5cm]{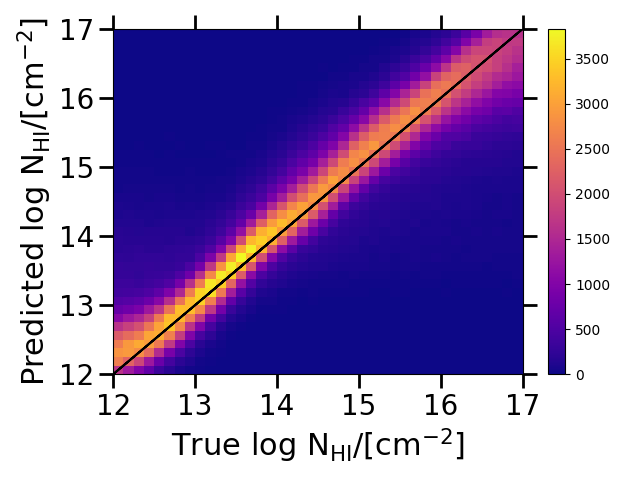}
    \includegraphics[width=6cm,height=5cm]{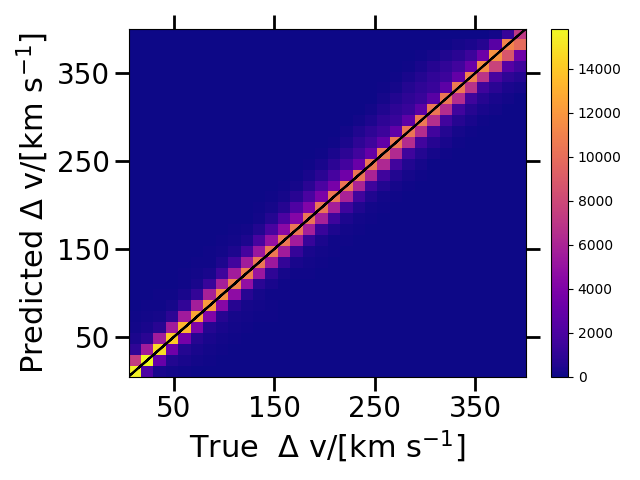}
    \includegraphics[width=0.31\textwidth,trim={0cm 0cm 0cm -1cm}]{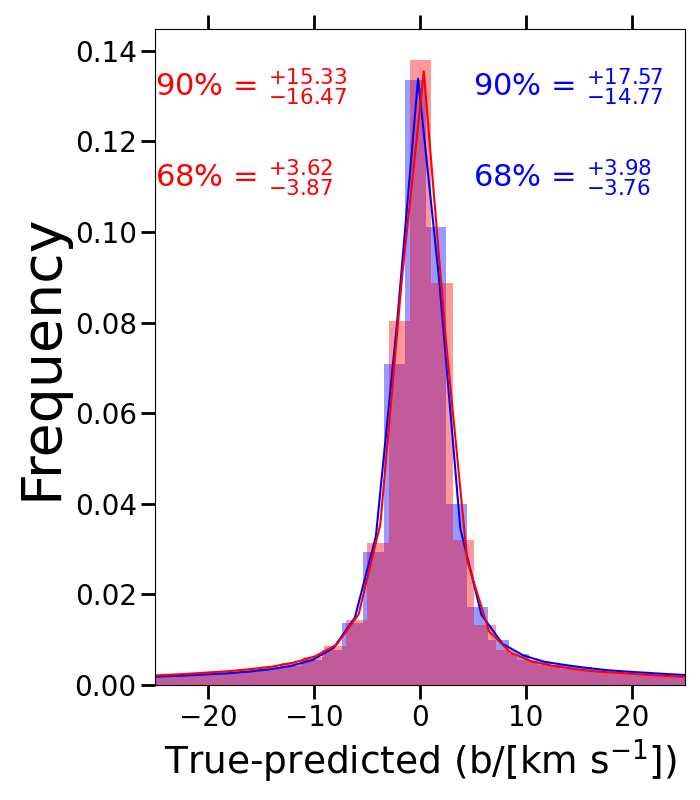}
    \includegraphics[width=0.31\textwidth]{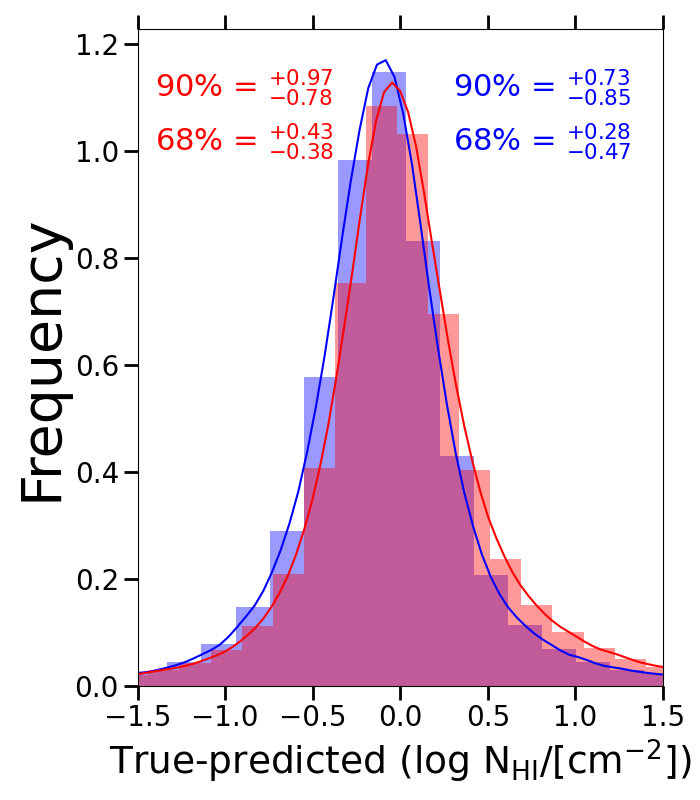}
    \includegraphics[width=0.31\textwidth]{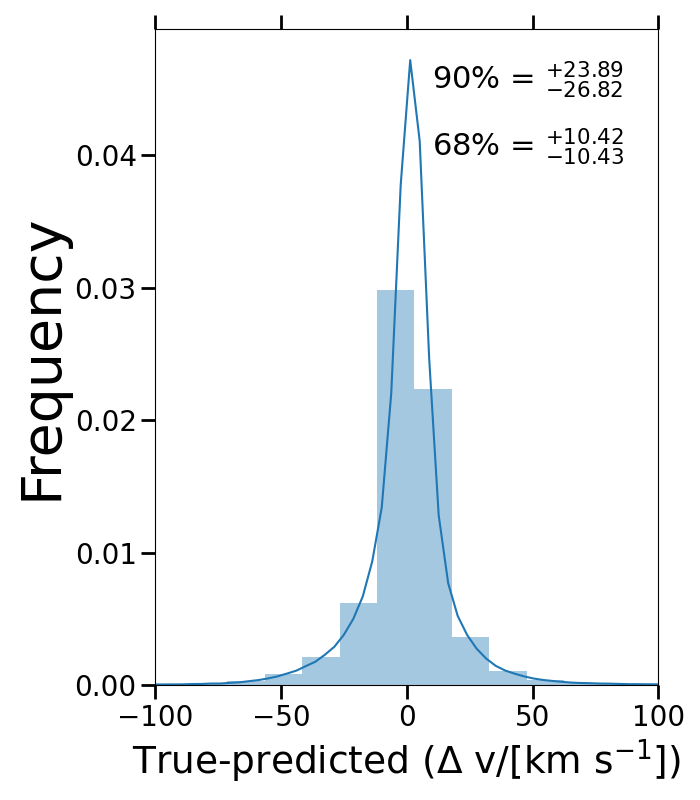}
 \includegraphics[width=6cm,height=5cm]{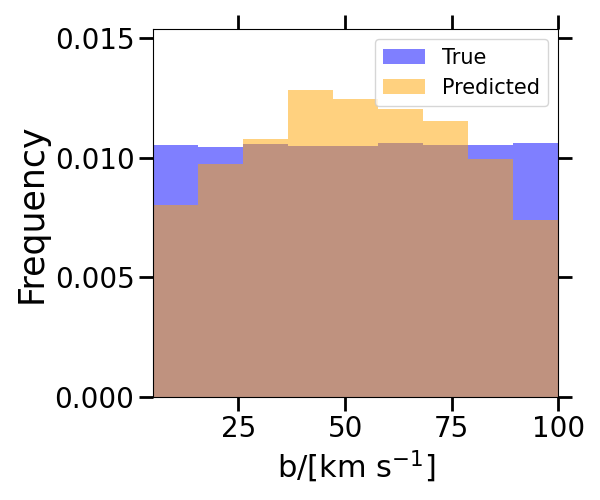}  \includegraphics[width=6cm,height=5cm]{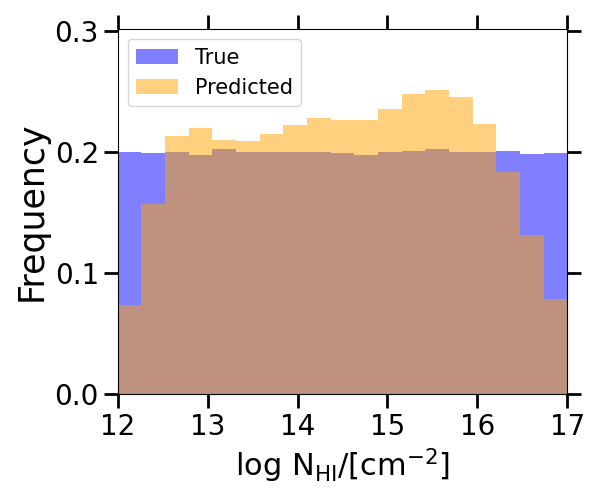}
    \includegraphics[width=6cm,height=5cm]{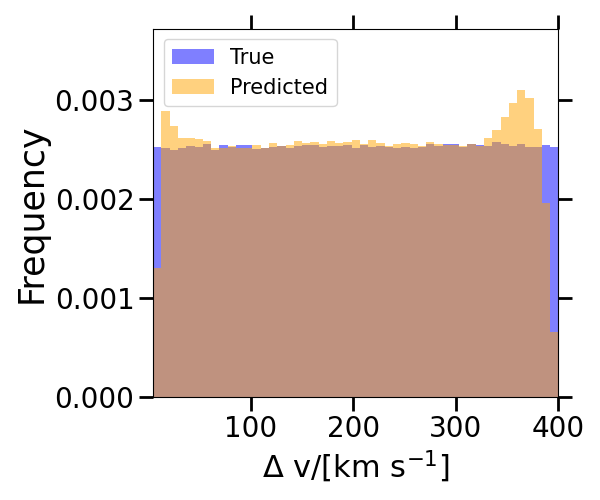}
    \caption{Same as Fig.~\ref{fig-result2} but for a double absorption line. The upper left panel shows the true versus predicted values stacking $b_1$  and $b_2$. Similarly, the upper middle panels show column density. The right upper panel shows the velocity difference between the two absorption lines. The middle panels show the histogram of the difference between true and predicted values, with 90\% and 68\% values marked at the top for component 1 (red color) and component 2 (blue color) of the double absorption line. The lower panel shows the normalized histogram of true and predicted parameters.}
    \label{fig-result3}
\end{figure*}

 \begin{figure}
     \centering
     \includegraphics[width=0.5\textwidth]{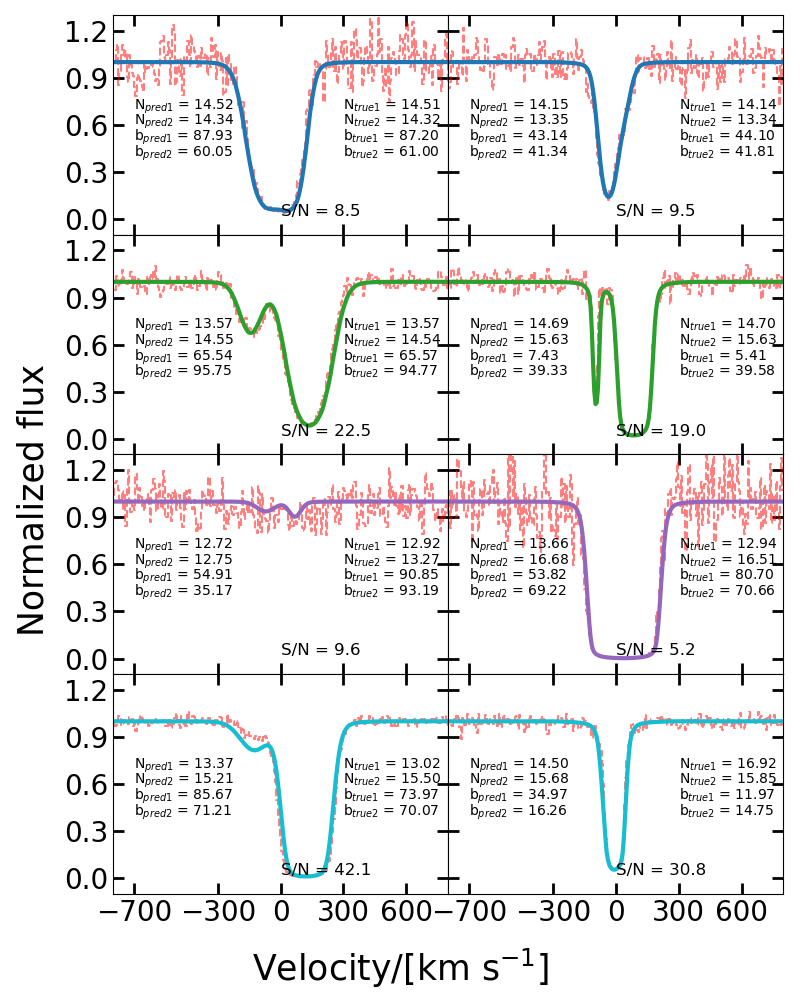}
      \caption{Same as Fig.~\ref{fig-cnn_n1_example} but examples of the simulated test double absorption lines (red-dashed line) and the corresponding Voigt profile predicted by the CNN model (in solid colored lines). The parameters $b$ and N$_\mathrm{HI}$ of two components (subscript 1 and 2) of the double line are written in each panel.}
     \label{fig-cnn_n2_example}
 \end{figure}

\subsection{Parameter estimation of the \lya absorption lines}
\label{sect-regr_result}
Figure~\ref{fig-result2} shows the performance of the CNN model predicting the physical properties of the simulated test dataset of a single \lya absorption line. The upper panels of Fig.~\ref{fig-result2} show the comparison between the true (horizontal axis) and predicted (vertical axis) values of Doppler width and column density for the test-simulated data. We find a tight correlation between the intrinsic and predicted parameters, indicating that our CNN model accurately predicts the physical properties of a single absorption line. The black line is the one-to-one line marking a perfect prediction. The figure shows a nominal scatter in the algorithm's prediction of $b$ and log N$_{\rm HI}$. We found negligible bias for the selected range of the parameters of $b$ and log N$_{\rm HI}$. 

The middle panels of Fig.~\ref{fig-result2} show the histogram of the difference between the true and predicted values of $b$ and log N$_{\rm HI}$. We find that 68\% [90\%] of the predictions for the Doppler width  ($\sigma_{68b}$ [$\sigma_{90b}$]) are within $^{+2.29}_{-1.06}$ [$^{+8.96}_{-6.68}$] km s$^{-1}$, and for column density (in log-scale, $\sigma_{68N}$ [$\sigma_{90N}$]) is within $^{+0.18}_{-0.16}$ [$^{+0.33}_{-0.35}$] cm$^{-2}$.  The combined ($\sigma_{90b}$ and $\sigma_{90N}$) outlier fraction with |b$_{true}$-b$_{pred}$| $>$ 7.8 \kms  along with |N$_{true}$-N$_{pred}$|  $>$ 10$^{0.33}$ cm$^{-2}$ is less than 0.01\%. This fraction reduces even further to only 0.0002\% if we consider output only for samples with S/N$>15$. The bottom panels of Fig.~\ref{fig-result2} show the histograms of true and predicted parameters. By construct, the count of true values is similar in each bin; however, CNN has relatively poor predictions of higher true $b$ values. This discrepancy suggests that the CNN may have difficulty accurately predicting higher $b$-values, potentially due to the complexity of spectral features associated with broad absorption lines. However, we do not observe similar evidence for column densities, indicating that CNN's predictions for column densities are relatively more accurate across different parameter ranges.

Figure~\ref{fig-cnn_n1_example} demonstrates a few examples of the predictions by CNN for the single absorption test-simulated data. The input data, consisting of 301 pixels generated in Sect.~\ref{sect-data}, is shown in red dashed lines. The figures in the upper boxes demonstrate precise physical parameter predictions made by the CNN for low S/N data (S/N$<10$; blue), while the second panel displays higher S/N data (S/N$>10$; green). The two lower panels illustrate two instances of an inaccurate prediction made by CNN for low (purple) and high S/N (cyan). 

The CNN results in similar accuracy for the test-simulated double absorption sample predictions as for the single absorption lines. The results comparing true and predicted values of $b$ ($b_1$ and $b_2$ ), log N$_{\rm HI}$ (N$_1$ and N$_2$) and $\Delta v$ are shown in the upper panel of  Fig.~\ref{fig-result3}. As seen from the upper panel of Fig.~\ref{fig-result3}, the density of predicted values overlaps the true versus true one-to-one black line. The scatter of predicted values of log N for double lines has increased compared to the scatter for single absorption lines. However, we do not see any biases in the parameter's predictions, even in double absorption lines. This alignment demonstrates the algorithm's robustness in handling these more complex absorption profiles. 

To test the algorithm's performance, we show the difference between the true and predicted values in the middle panel of Fig.~\ref{fig-result3}. The histogram shows that $\sigma_{68b}$ [$\sigma_{90b}$] for $b_1$ is within $^{+3.62}_{-3.87}$ [ $^{+15.33}_{-16.47}$] km s$^{-1}$, and that for $b_2$ is within $^{+3.98}_{-3.76}$ [$^{+17.57}_{-14.77}$] km s$^{-1}$. Similarly $\sigma_{68N}$ [$\sigma_{90N}$] for log N$_1$  is within $^{+0.43}_{-0.38}$ [$^{+0.97}_{-0.78}$] cm$^{-2}$ and that for log N$_2$  is within $^{+0.28}_{-0.47}$ [ $^{+0.73}_{-0.85}$] cm$^{-2}$ and that for $\Delta v$ is $^{+10.42}_{-10.43}$ [$^{+23.89}_{-26.82}$] km s$^{-1}$. The combined ($\sigma_{90b}$ and $\sigma_{90N}$) outlier fraction with |b$_{true}$-b$_{pred}$| $>$ 16 \kms  along with |N$_{true}$-N$_{pred}$|  $>$ 10$^{0.90}$ cm$^{-2}$ is less than 2.5\%. This fraction reduces even further to only 1.5\% if we consider output only for samples with S/N$>15$. 
{The lower panel of Fig.~\ref{fig-result3} illustrates that CNN predictions are less accurate for lines with extreme $b$-values, log N$_{\rm HI}$, and $\Delta v$. When $b$ and N$_{\rm HI}$ values are low, accuracy decreases due to the signal being obscured by noise. Conversely, for higher $b$ and lower $\Delta v$ values, lines tend to blend together, leading to less accurate predictions. Similarly, high column densities result in poor predictions due to saturation effects. For higher $\Delta v$ values, the predictions are affected because the two lines become nearly separate entities. In this scenario, either one line may be obscured by noise, or both lines may be nearly saturated, impacting the accuracy of predictions.

A few examples of the double absorption test-simulated data with their predicted Voigt profiles are shown in Fig.~\ref{fig-cnn_n2_example}. Notably, our ML model can accurately predict two parameters even if the double absorption lines are separated by small $\Delta v$. However, challenges arise when either the absorption lines are nearly saturated, or they are heavily obscured by noise, or there is a significant contrast in optical depth between two absorption lines with one almost buried in noise. These scenarios emphasize the CNN model's limitations under specific circumstances and scope for improvement.

Based on this analysis, we conclude that our CNN models accurately predict the column density and Doppler width for single and double absorption lines. The minimal outliers confirm that CNN can robustly extract $b$ and N$_{\rm HI}$ of the single and double absorption lines for the simulated test data.

\section{Application to real data}
\label{sect-real}
The preceding sections show that our ML algorithm excels in its performance on simulated absorption lines designed to emulate the characteristics of real data obtained from the HST-COS, as detailed in Sect.~\ref{sect-data}. To evaluate the real-world applicability of our ML algorithm (Sect.~\ref{sect-class} and Sect.~\ref{sect-regr}), we now test it on \lya line profiles obtained from the COS data \citepalias{Danforth2016}. To ensure compatibility, we chose observed data falling within a parameter range akin to the simulated data (as outlined in Section \ref{sect-data}), in addition to 3$\sigma$ detection according to the line detection criteria used in \citetalias{Danforth2016}, resulting in 1364 absorption systems.

The line list files provided by \citetalias{Danforth2016}\footnote{available at https://archive.stsci.edu/prepds/igm/} include 2400 \lya absorption lines. We examine the corresponding spectrum for each line to determine whether the \lya line consists of single or multiple components. Our approach involves searching each line to identify if adjacent lines share common wavelengths. If the span of common wavelengths for absorption lines continuously touches the spectrum's continuum (within error of the spectrum) for more than seven pixels (with each pixel being 0.035 \AA), we classify these as single-component absorption lines. For the remaining lines, we count the number of lines meeting the aforementioned criteria. With this method, we find that out of 2400 components, approximately 1557 are \lya lines with a single component, and 286 systems are identified as having double components. Following specific criteria ($5 \leq b/$\kms$ \leq 100$, $12 \leq$ log N/cm$^{-2} \leq 17$, $5 \leq$ S/N $\leq 100$ and discarding lines with significance below 3$\sigma$), we selected 1364 lines for further analysis.

To use these absorption lines as inputs for our ML algorithms, we select a segment of the spectrum encompassing the absorption line and an adjacent line-free region. This segment is centered within a 301-pixel chunk. If the spectral segment does not span the entire 301 pixels, we pad the remaining locations with a continuum added with Gaussian random noise. The standard deviation of the noise in this padded area is determined based on the median of the error vector within the line-free region surrounding the absorption line. To assess the performance of our ML algorithms on this dataset, we first apply our ML algorithm to these lines and record the results. For comparison purposes, we employ two distinct Voigt profile fitting algorithms. In addition to fitting done by \citetalias{Danforth2016}, we also apply the VIPER algorithm (\citetalias{Prakash2017}) to the same dataset. 

Before analyzing the performance of our algorithms, it is important to highlight the differences in determining the number of components in absorption line systems between \citetalias{Danforth2016} and VIPER. Although VIPER adheres to the same significance level criteria for line identification as \citetalias{Danforth2016}, their methods for determining whether lines are single or multicomponent differ. \citetalias{Danforth2016} used additional spectral information, such as higher-order lines or coincidental metal lines, to determine if lines are single or multicomponent. In contrast, VIPER employs the Akaike Information Criterion with Corrections (AICC) to decide the number of components. Furthermore, if AICC identifies more than one component, VIPER re-evaluates the significance level of all components, discarding any component with significance below $3 \sigma$. Due to these methodological variations, VIPER identifies 784 single lines and 296 double lines in the sample, differing from the 1137 single and 227 double lines found by \citetalias{Danforth2016}. Given that VIPER, like our ML algorithm, does not utilize additional spectral information for component identifications and relies solely on \lya lines as input, we anticipate our algorithm to exhibit improved performance in classification when using VIPER labels as compared to \citetalias{Danforth2016} labels.

\subsection{Performance on HST COS data: Classification}
First, we use the classification algorithm (see Sect.~\ref{sect-class_model}) to test the accuracy in predicting the number of absorption lines for the COS data. Similar to Sect.~\ref{sect-class_model}, we input the normalized flux to the model, and the output is the number of absorption lines. We visualize the model's performance by comparing the predicted labels with ``true'' labels obtained from the fits of  \citetalias{Danforth2016} and VIPER. 

In Fig.~\ref{fig-result2_real_matrix}, we compare the results of our classification algorithm with the results of \citetalias{Danforth2016} (the left-hand panel) and VIPER (right-hand panel). In comparison with \citetalias{Danforth2016}, we find that the double lines are identified accurately (93.4\%). However, there are 22\% of single lines identified as double lines, resulting in the accuracy for single lines to be just 78\%. Upon further investigation, we found that most of the 22\% single lines (152 out of 255) that are identified as double lines have S/N $<$ 20. The overall accuracy of our algorithm with respect to the classification of single versus double lines identified by \citetalias{Danforth2016} is 80.21\%, whereas other metrics such as recall is 77.57\%, F1-score is 0.86.

\begin{figure}[!t]
    \centering
    \includegraphics[width=4cm,height=4cm]{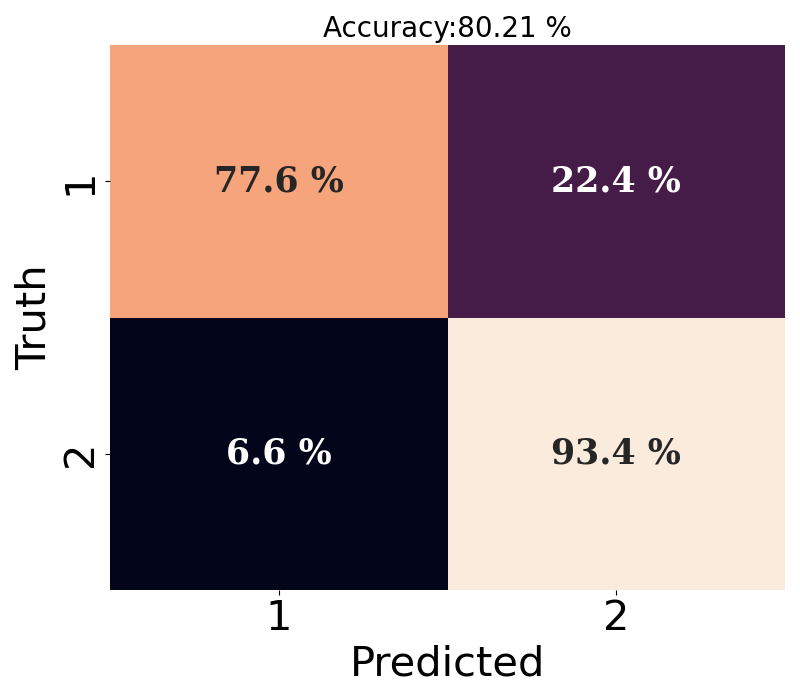} 
    \includegraphics[width=4cm,height=4cm]{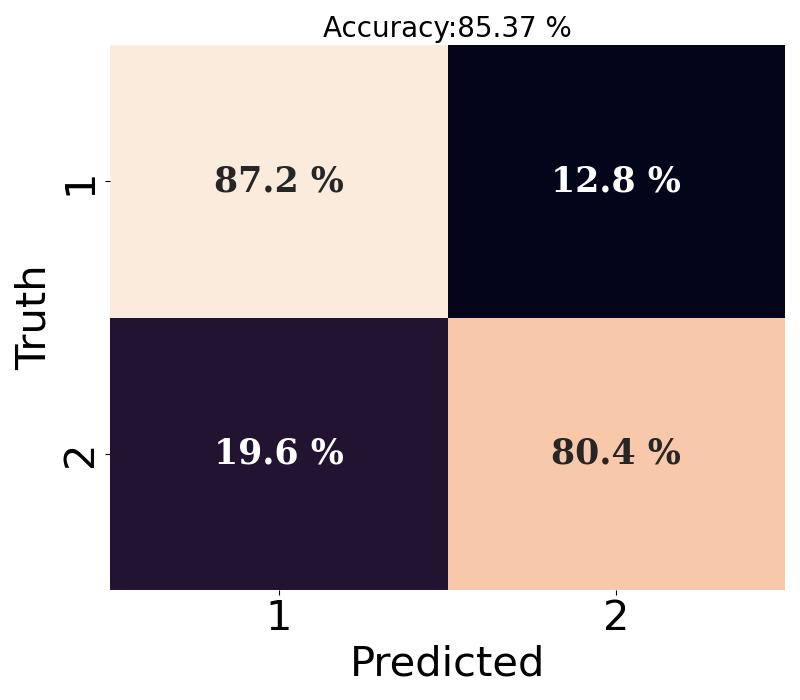} 
    \caption{The plot shows the prediction from the classification algorithm for the HST data with true labels from \citetalias{Danforth2016} (left) and VIPER (right). For \citetalias{Danforth2016} [VIPER], we find the accuracy = 80.21\% [85.37\%],  sensitivity=77.57\% [87.24\%], specificity=93.39\% [80.41\%], precision=98.33\% [92.18\%] and negative predictive value=45.40\% [70.41\%].}
    \label{fig-result2_real_matrix}
\end{figure}

\begin{figure}[!t]
    \centering
    \includegraphics[width=4cm,height=4cm]{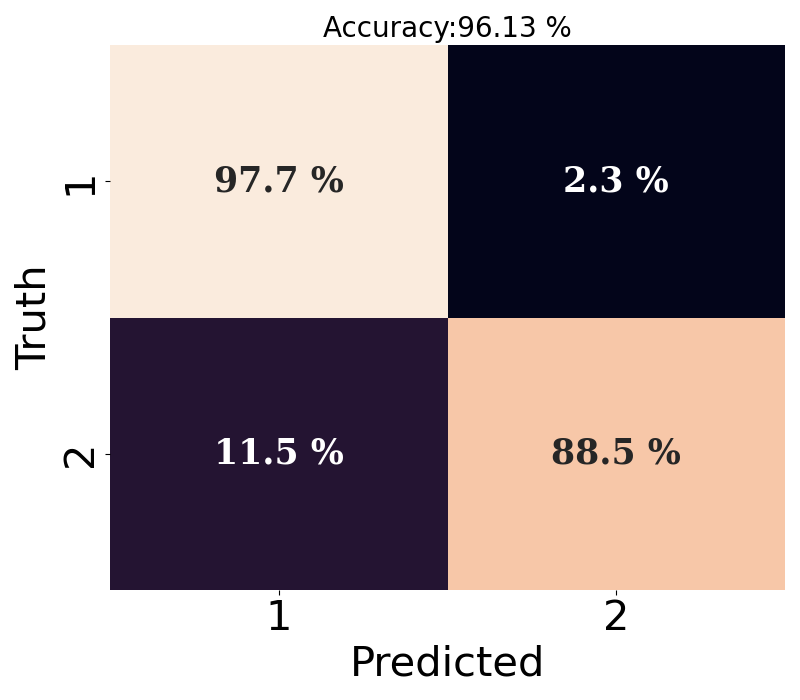} 
    \includegraphics[width=4cm,height=4cm]{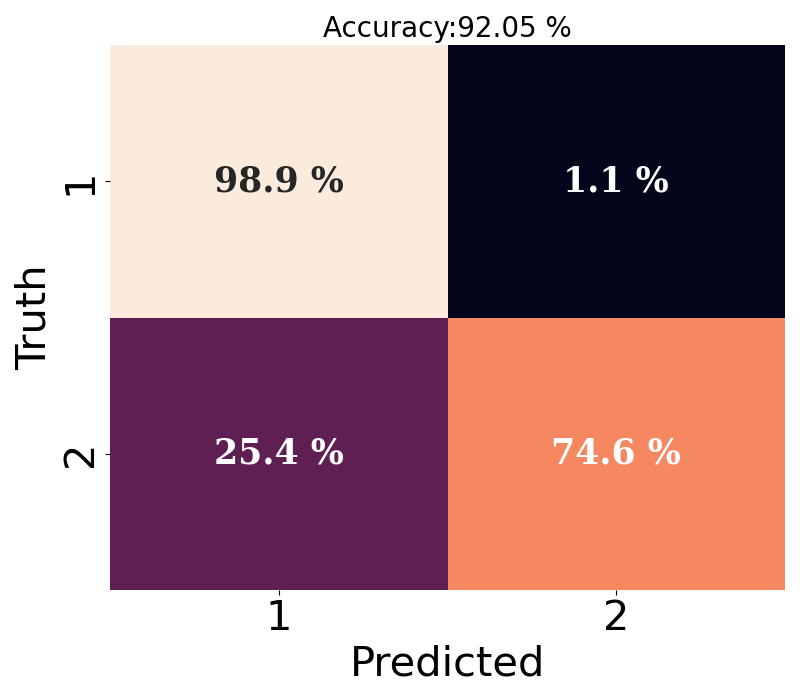} 
    \caption{Same as Fig.~\ref{fig-result2_real_matrix} but for the mock dataset. For \citetalias{Danforth2016} [VIPER], we find the accuracy = 96.13\% [92.05\%], sensitivity=97.71\% [98.94\%], specificity=88.50\% [74.58\%], precision=97.62\% [90.79\%], and negative predictive value=88.89\% [96.54\%].}
    \label{fig-result2_mock_matrix}
\end{figure}

\begin{figure*}
    \centering
    \includegraphics[width=0.46\textwidth]{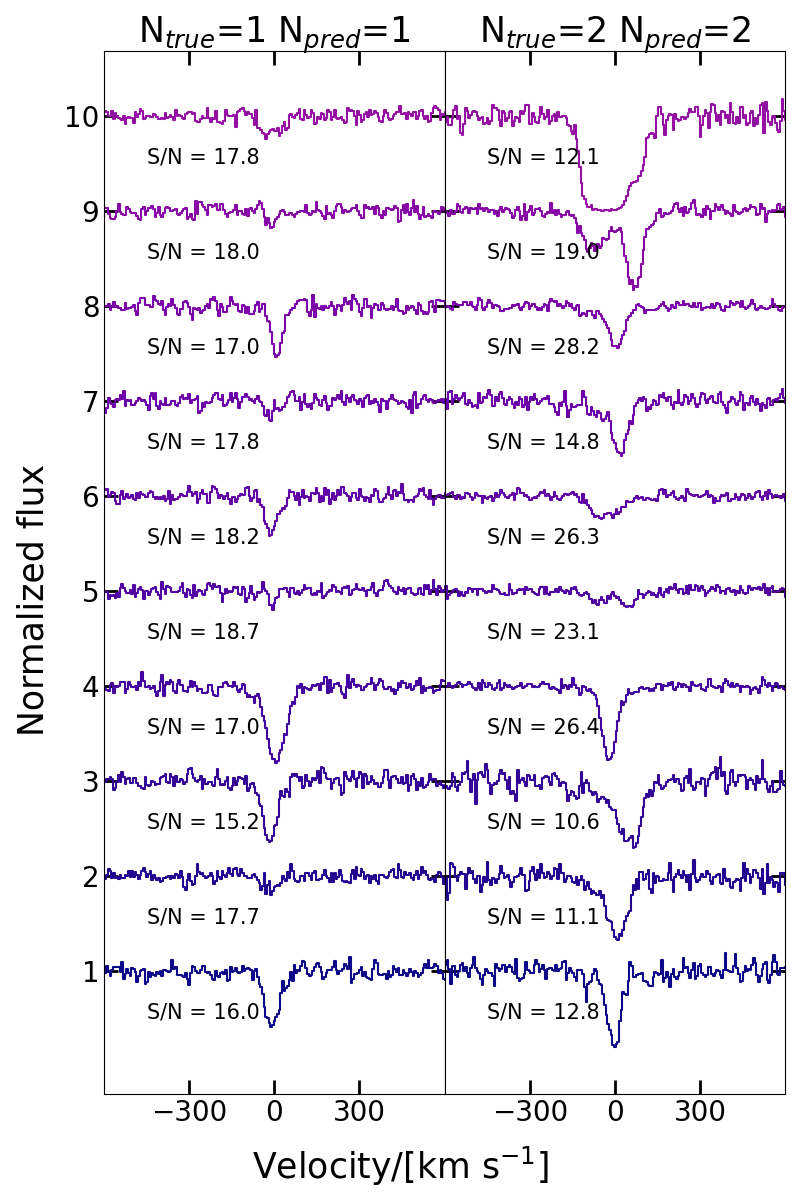}  
    \includegraphics[width=0.46\textwidth]{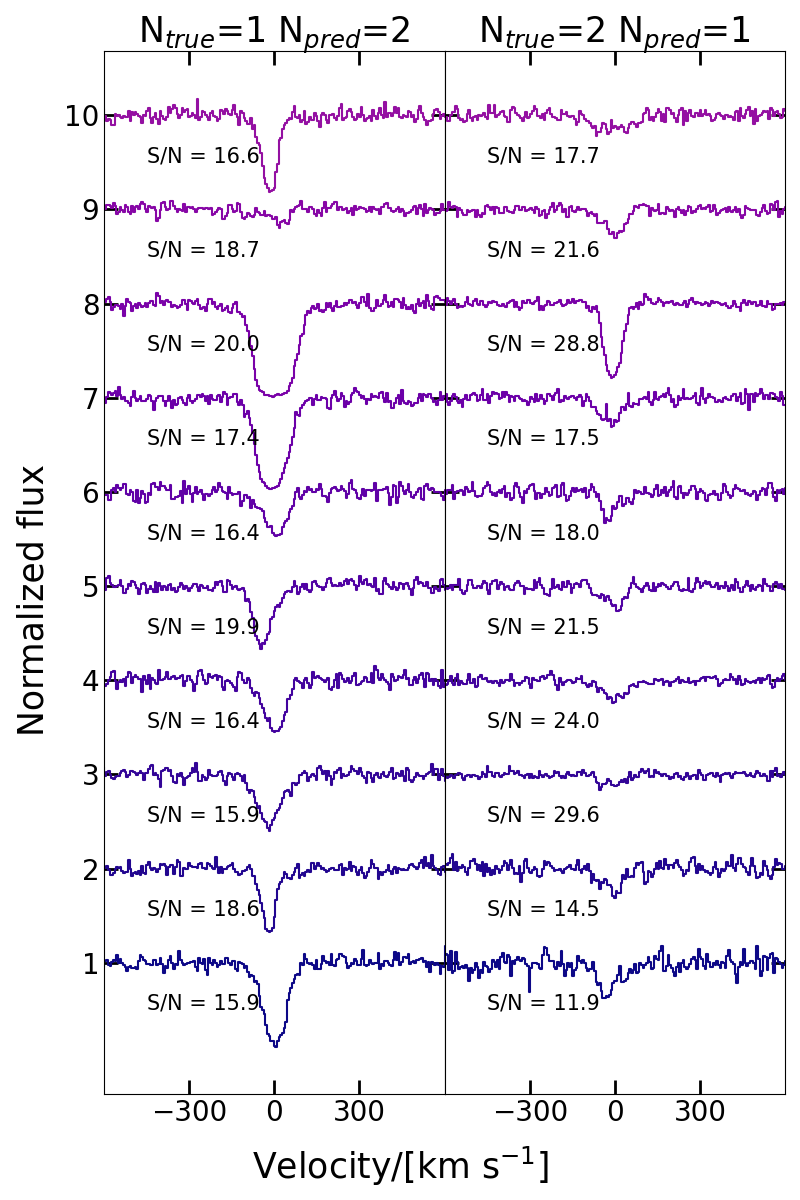}
     \caption{Examples of classification performance of real observed absorption Lines. {\it Left Panel}: Successful Classification - The panel displays instances where the classification algorithm accurately identifies single and double absorption lines, effectively matching the true labels. {\it Right Panel}: Misclassified Cases - The two panels show examples of the misclassified single and double absorption lines, highlighting areas for improvement in certain challenging scenarios.}
    \label{fig-class_miss}
\end{figure*}

\begin{figure*}
    \centering
    \includegraphics[width=0.45\textwidth]{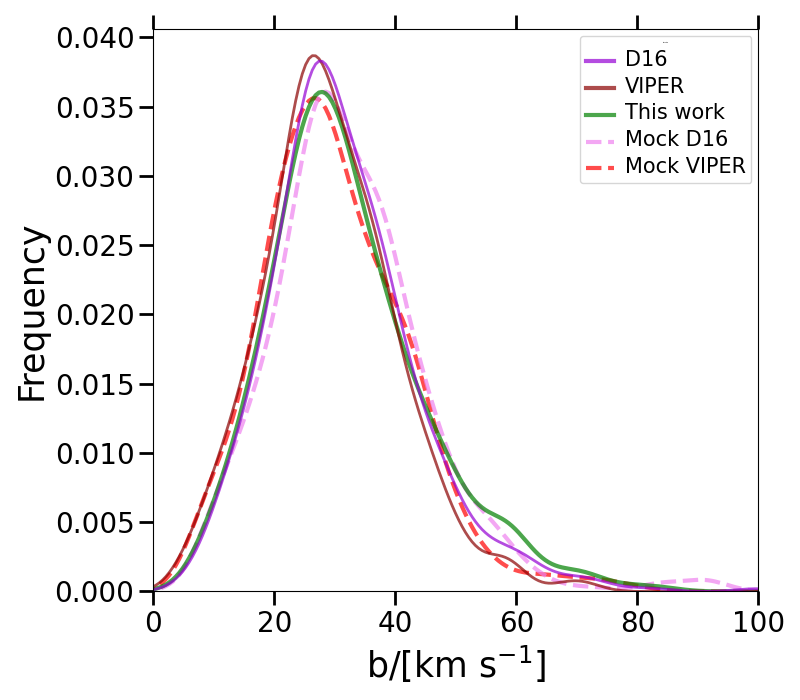}
    \includegraphics[width=0.45\textwidth]{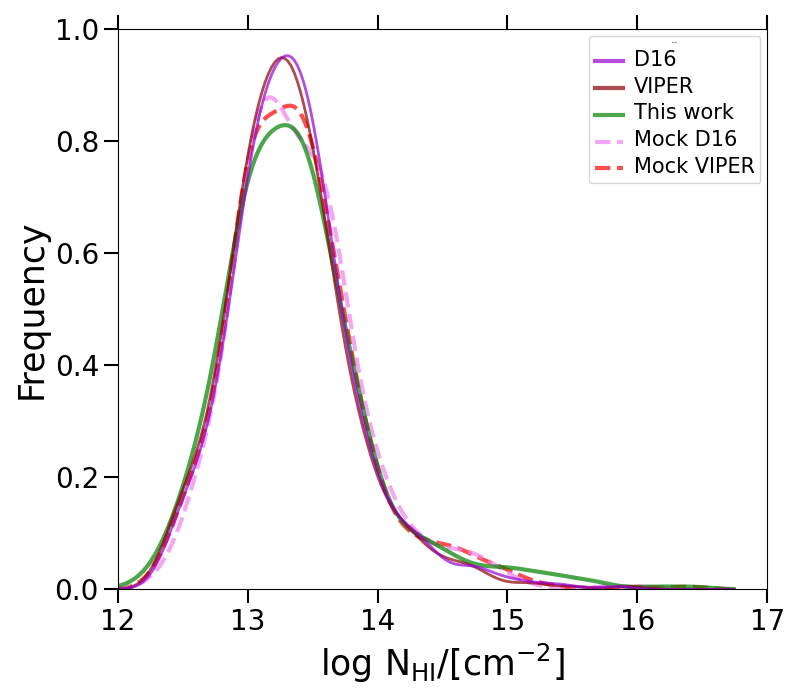}   
   \includegraphics[width=0.23\textwidth]{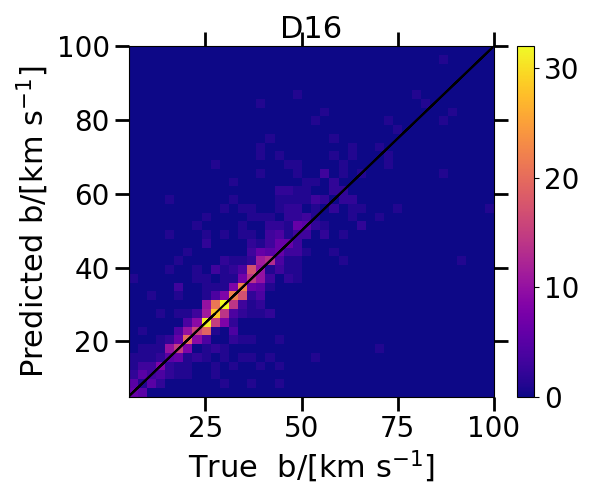}
   \includegraphics[width=0.23\textwidth]{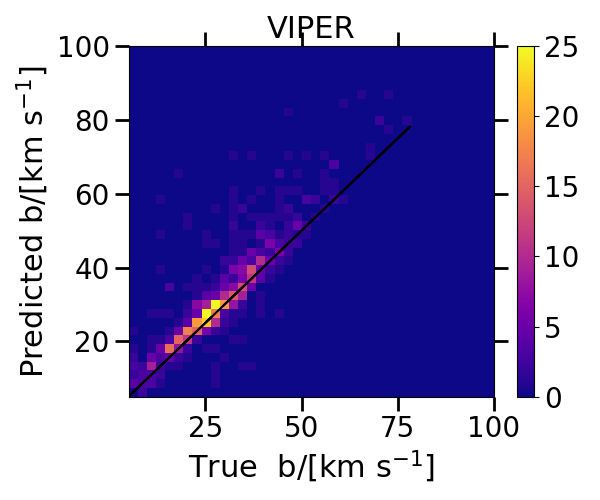}
   \includegraphics[width=0.23\textwidth]{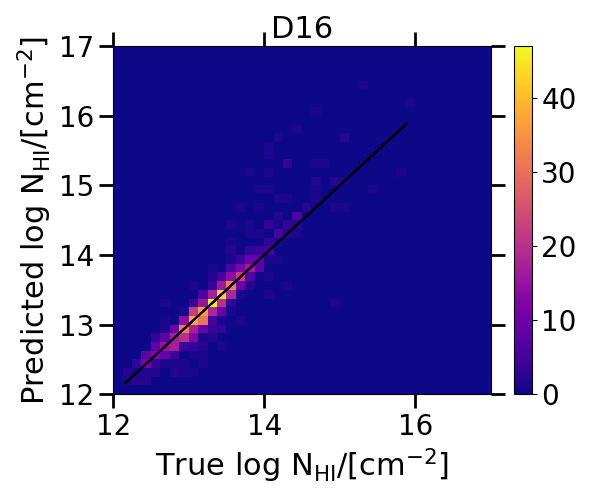}   
   \includegraphics[width=0.23\textwidth]{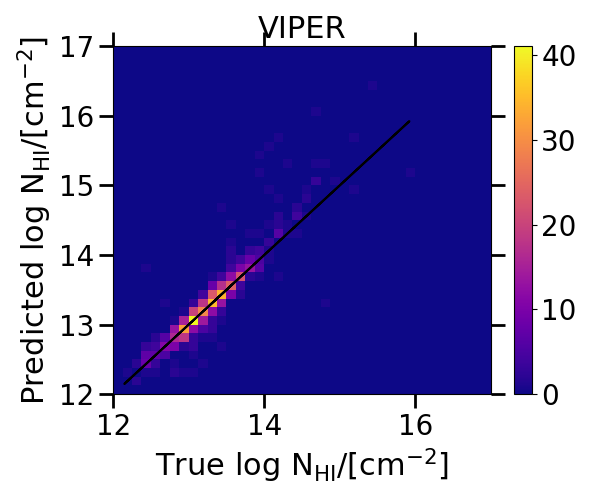}   
     \caption{Comparison of true parameter distributions and predictions for real observed single absorption lines. {\it Upper panels:} The histograms show the distribution of $b$ and logN$_{\rm HI}$ for a single absorption line common in all three studies estimated by the CNN model in this work (green) overplotted with distributions from \citepalias{Danforth2016} (purple) and  VIPER (red). The distributions of the CNN predictions for the corresponding mocks are shown in dashed lines.{\it Lower panels:} The prediction of single line parameters from CNN compared to the true values from \citepalias{Danforth2016} and VIPER.}
    \label{fig-result2_real1}
\end{figure*}

\begin{figure*}
    \centering
    \includegraphics[width=0.45\textwidth]{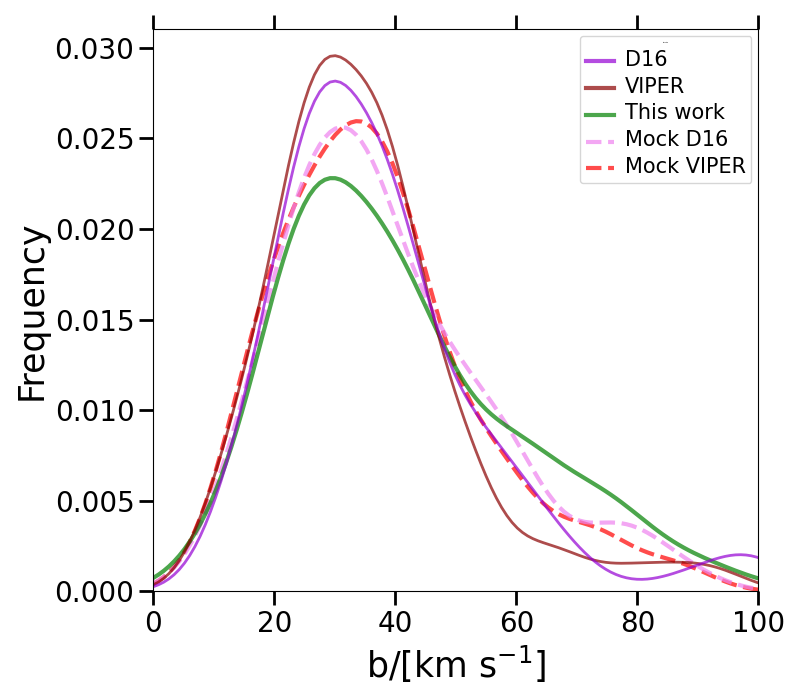}
    \includegraphics[width=0.45\textwidth]{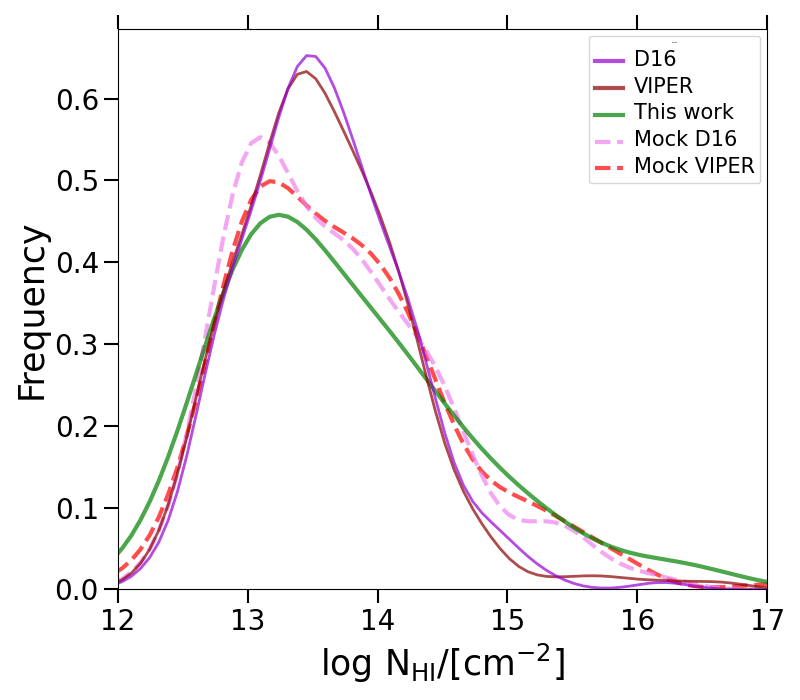} 
    \includegraphics[width=0.22\textwidth]{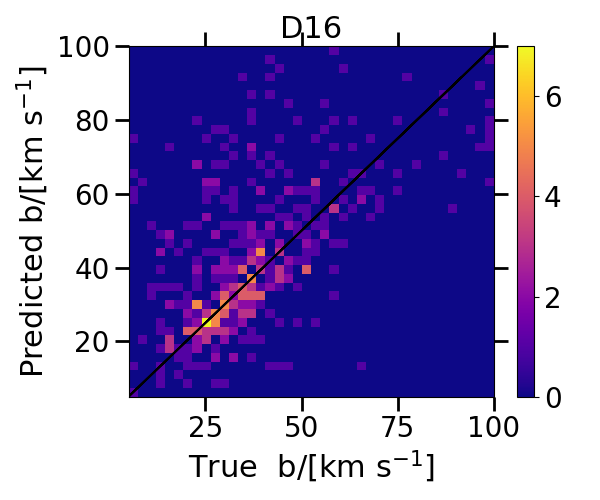}
    \includegraphics[width=0.22\textwidth]{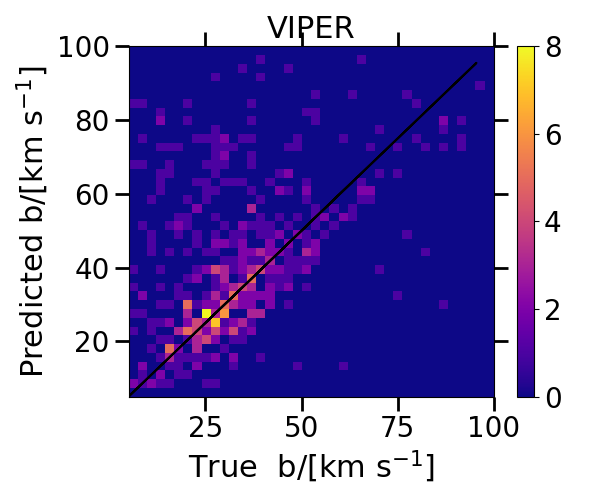}
    \includegraphics[width=0.22\textwidth]{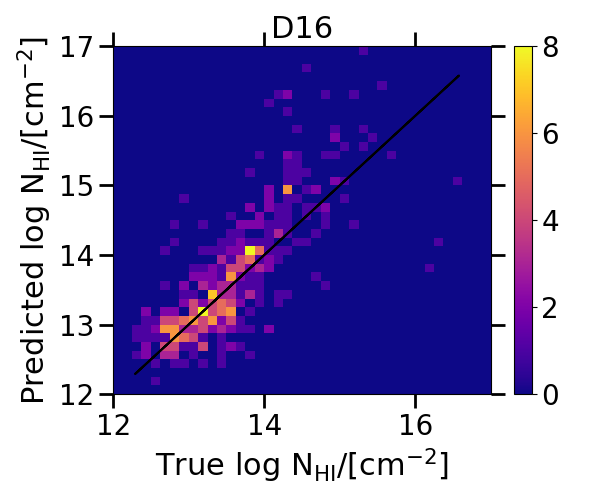} 
    \includegraphics[width=0.22\textwidth]{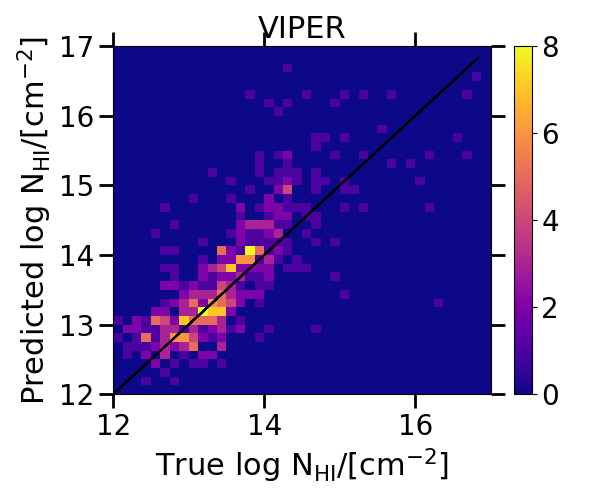} 
    \caption{Same as Fig.~\ref{fig-result2_real1} but for parameters extracted from double absorption lines.}
    \label{fig-result2_real2}
\end{figure*}
\begin{figure*}
    \centering
    \includegraphics[width=0.45\textwidth]{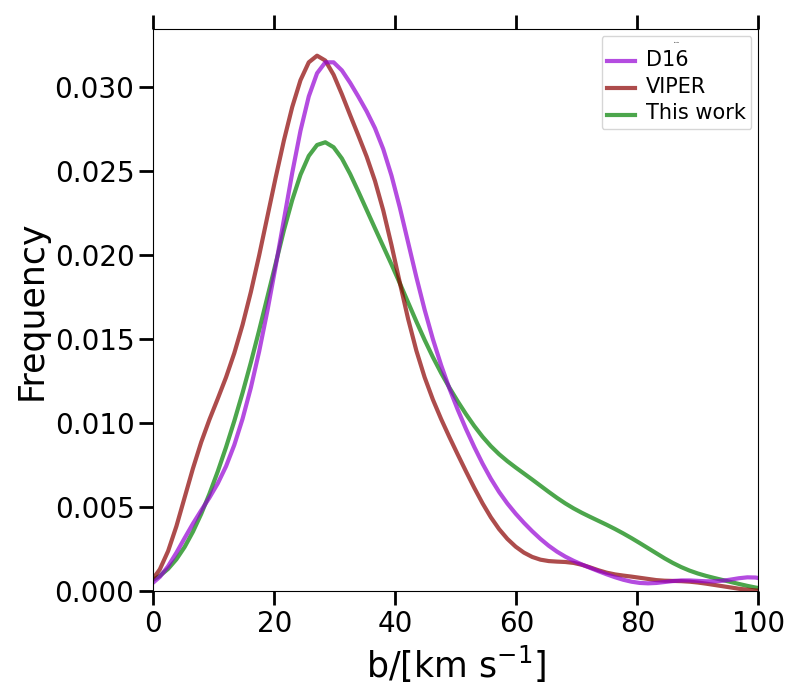}
    \includegraphics[width=0.45\textwidth]{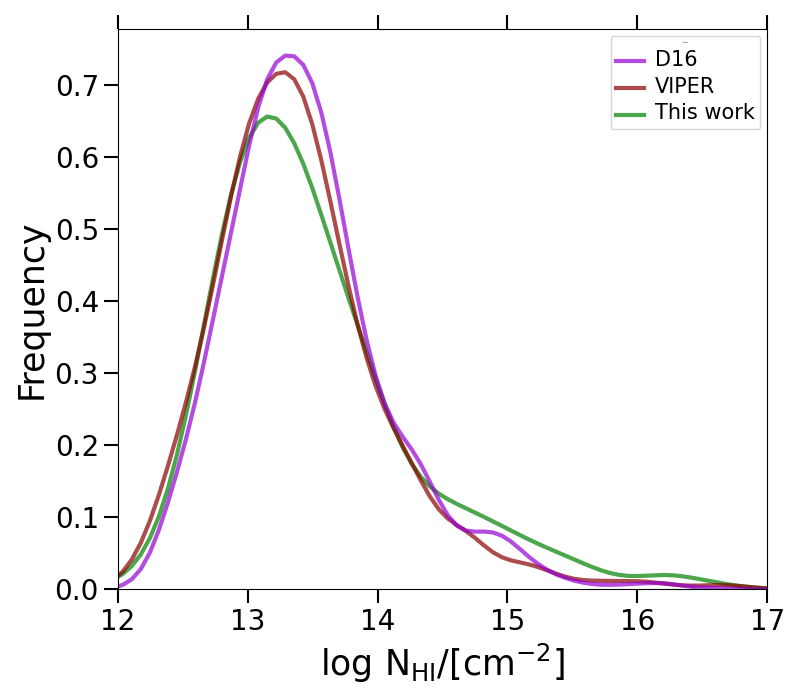} 
    \caption{ $b$ and log N$_{\rm HI}$ extracted from all the absorption lines. All the lines identified as single lines by our classification model are inputted into regression model 1, and all the lines identified as double lines by our classification model are inputted into regression model 2. The output from both of these models is plotted as the green line in comparison to the $b$ and log N$_{\rm HI}$ estimates by \citetalias{Danforth2016} (purple) and VIPER (red) extracted from their single and double identified lines.}
    \label{fig-result2_real3}
\end{figure*}

In comparison with the results of VIPER (see the right-hand panel in Fig.~\ref{fig-result2_real_matrix}), we find that 87\% of the single lines are identified correctly, and only 13\% were misclassified as double lines. For double lines, however, our algorithm could classify correctly for 80\% of the lines and miss-classify 20\% of the lines as single lines. 47 out of 58 misclassified double lines have S/N$<20$. The overall accuracy of our algorithm with respect to the classification of single versus double lines identified by VIPER is 85.37\%, which is better than the one obtained for identification \citetalias{Danforth2016}, as per our expectations. Other metrics, such as sensitivity (recall), are 87.24\%, and F1-score is 0.89. 

Although our algorithm performs reasonably well, achieving an accuracy of 85\% for labels from 
VIPER, this does not match the 93\% accuracy obtained with our simulated data (refer to Fig.~\ref{fig-confusion_matrix}). The 8\% decrease in accuracy is primarily due to a 10\% reduction in true positives and true negatives for real data. We conducted a visual inspection of several instances where our algorithm was unsuccessful, yet we could not identify a clear reason for these failures. In many cases, the misclassifications appeared to stem from genuine confusion where double lines look single or are not prominent because of poor S/N. Examples of such cases, including those where the classification was accurate, are depicted in Fig.~\ref{fig-class_miss}. 

A potential reason for an 8-10\% decrement in the performance of our algorithm on real data as compared to simulated data might be the inherent difficulty in emulating the real observations. To test this hypothesis, we decided to utilize our mock dataset (see Sect.~\ref{sect-mock}) where we used exact same parameters of both \citetalias{Danforth2016} and VIPER fits, that is, identifications as well as $b$ and N$_{\rm HI}$ values and modeled instrument effects, noises, and central wavelength. We then input those in our classification algorithm and compared them with \citetalias{Danforth2016} and VIPER labels. In this case, our results are shown in Fig.~\ref{fig-result2_mock_matrix}. We find that the accuracy [F1-score] for \citetalias{Danforth2016} fits is  96.13\% [.97], and VIPER fits is 92.05\% [.94]. These accuracies are comparable to the accuracy obtained in the simulated data. Therefore, it seems our hypothesis is correct, and there are some subtle differences between emulating the real data and the real data. Noise in real data originates from various sources, including detector noise and imperfections, cosmic rays, and photon noise.  These noise patterns are inherently complex and can fluctuate over time and across different wavelengths. In contrast, simulated data incorporate noise patterns generated through simplified or idealized models and may not perfectly match the characteristics of real noise. This difference is even more apparent for regression analysis, as discussed in the following subsection.

\subsection{Performance on HST COS data: Regression }

After testing the classification of absorption lines, we now analyze the performance of the regression model on the identified absorption lines. This analysis entails utilizing flux data from \citetalias{Danforth2016} specifically for lines categorized as single (784) and double (296) by VIPER, totaling 1080 lines. Given the greater consistency observed in our classification algorithm with VIPER, we opt to utilize these line predictions. Among the 784 single lines, our classification model identified 684 as single lines, and among 296 double lines, our classification model identified 238 correctly.

First, we input the 620 (out of 684) single lines common across all studies (\citetalias{Danforth2016}, VIPER and this work) into our algorithm to predict the values of $b$ and N$_{\rm HI}$. In the upper panel of Fig.~\ref{fig-result2_real1}, we illustrate the distribution of $b$ and N$_{\rm HI}$ from five datasets: \citetalias{Danforth2016} (purple lines), VIPER (red lines), CNN estimates of the real data (green lines), CNN estimates of \citetalias{Danforth2016} mocks (purple dashed lines), and CNN estimates of VIPER mocks (red dashed lines). We find a good consistency between the various studies. From our analysis of mock data, we discovered that centering the absorption lines significantly influences the accuracy of estimates. Therefore, all the absorption lines were centered before input into the ML model.

Our classification algorithm successfully predicted 882 single lines out of 1137 from the \citetalias{Danforth2016} dataset and 684 single lines out of 784 from the VIPER dataset. In the lower panel of Figure~\ref{fig-result2_real1}, we compare the CNN-predicted  $b$ and N$_{\rm HI}$ of these predicted single lines with the actual labels from \citetalias{Danforth2016} (882) and VIPER (684). The true versus true one-to-one line is overlaid in black for comparison. Interestingly, while classification results were more consistent with VIPER as compared to \citetalias{Danforth2016}, this consistency does not extend to parameter estimation results. This discrepancy may stem from differences in how higher-order lines are utilized for classification in \citetalias{Danforth2016}. However, fitting techniques remain independent of the use of higher-order lines in all the studies.

{The similar inconsistencies between CNN prediction and \citetalias{Danforth2016} or VIPER is more clearly evident from  Table~\ref{tab:inco}. Table~\ref{tab:inco} provides details on the fraction of data exhibiting inconsistencies between the true (Data) and CNN-predicted estimates using $\sigma_{90b}$ and $\sigma_{90N}$ defined in Sect.~\ref{sect-regr_result}. We mention the fraction of sample with  $\Delta b = b_{\text{true}}$ - $b_{\text{pred}}$ $> \sigma_{90b}$ or $\Delta N = N_{\text{true}}$ - $N_{\text{pred}}$ $> \sigma_{90N}$. In our analysis of simulated test data, we anticipate that the outcomes may vary depending on the S/N. Therefore, we examined the disparities of physical parameters between true and CNN predictions for the two S/N bins, that is, S/N$<20$ and S/N$>20$.

For the single lines, rows 1 and 3 of the table show the discrepancy between the CNN predictions for real data in comparison to the true parameters estimated from \citetalias{Danforth2016} and VIPER. Rows 2 and 4 show the same but for CNN predictions of the corresponding mock data. Notably, the mock tests exhibit reduced misestimations compared to real data, underscoring the challenge of accurately simulating real observations. Reiterating the fact that the real data are already centered before input to the CNN model, the better consistency between true and predicted values for mock datasets underscores the challenge of simulating real observations accurately. Nevertheless, these mock tests demonstrate the successful prediction of $b$ and N$_{\rm HI}$ parameters by our ML algorithm, particularly when the simulated training dataset closely resembles the test dataset.

We observed that for S/N$>20$, the results demonstrate greater consistency for column density, with a maximum of 2.7\% [0.5\%] of cases where predictions from the CNN do not align within $\sigma_{90N}$ for the real [mock] dataset. However, for the prediction of $b$, we find similar results for S/N$<20$ or S/N$>20$.  We find almost no absorption line at S/N$>20$ that has both $b$ and N$_{\rm HI}$ to be uncertain. Additionally, as indicated in Table~\ref{tab:inco}, we observe that most of the inconsistencies arise from smaller values of $b$ and N$_{\rm HI}$, particularly at higher S/N levels.

We extend our evaluation of the ML algorithm to encompass 118 (out of 238) common double absorption lines, which involve additional physical parameters including $b_1$, $b_2$, N$_1$, N$_2$, and $\Delta v$. In the upper-left panel of Fig.~\ref{fig-result2_real2}, we present stacked values of $b_1$ and $b_2$, while the upper-right panel illustrates stacked values of log N$_1$ and log N$_2$. Notably, our ML algorithm predicts $b$ values significantly higher than those estimated by both \citetalias{Danforth2016} and VIPER. Similarly, the N$_{\rm HI}$ estimates generated by our ML algorithm consistently surpass those obtained by \citetalias{Danforth2016} and VIPER. Similar to single lines, the S/N and centering of double lines significantly impact the results.  This impact is evident from our analysis of mock dashed lines.  However, centering double absorption lines poses additional challenges and hence is not performed.

Our classification algorithm successfully predicted 212 double lines out of 227 from the \citetalias{Danforth2016} dataset and 238 double lines out of 296 from the VIPER dataset. In the lower panel of Fig.~\ref{fig-result2_real2}, we compare CNN predicted $b$ and N$_{\rm HI}$ of these double lines with true estimates from \citetalias{Danforth2016}  and VIPER. Interestingly, the scatter of predicted values of log N$_{\rm HI}$ for double lines increases compared to single absorption lines, consistent with results from simulated test data  (Fig.~\ref{fig-result3}). This suggests that predictions for double lines are less accurate than those for single lines.

\begin{table*}[htbp]
  \centering
  \caption{ Summary of the percentage of absorption lines with a discrepancy between true physical parameters and CNN predictions exceeding $\sigma_{90}$ (90\% percentile derived from simulated test data). $\sigma_{90b}$ and $\sigma_{90N}$ for single line is taken from Fig.~\ref{fig-result2} and for double line from  Fig.~\ref{fig-result3}. Single-line data are shown in rows 1-4, double-line component 1 (C1) in rows 5-8, and double-line component 2 (C2) in rows 9-12. }
  \begin{tabular}{lll rr rr cc}
    \hline
 \multirow{2}{*}{Sno.} &   \multirow{2}{*}{Lines} & \multirow{2}{*}{Data} 
        & \multicolumn{2}{c}{$\Delta b > \mathrm{\sigma_{90b}} [\mathrm{median}\, N]$} & \multicolumn{2}{c}{$\Delta N > \mathrm{\sigma_{90N}} [\mathrm{median}\, b]$}  & \multicolumn{2}{c}{$\Delta b > \mathrm{\sigma_{90b}} \& \Delta N > \mathrm{\sigma_{90N}}$}   \\
    \cmidrule(rr){4-5}   \cmidrule(rr){6-7} \cmidrule(rr){8-9} 
    &  &  & S/N$<$20 &  S/N$>$20 &  S/N$<$20  &  S/N$>$20&  S/N$<$20  &  S/N$>$20          \\
    \midrule
1 & Single & D16 & 12.8\% [13.35] & 14.8\% [12.70] & 9.4\% [32.50] & 2.4\% [34.60] & 5.0\% & 0.3\% \\ 
2 & Single & D16 mock & 6.2\% [13.24] & 7.5\% [12.77] & 2.8\% [28.70] & 0.5\% [18.80] & 0.4\% & 0.0\% \\ 
3 & Single & VIPER & 14.1\% [13.22] & 13.0\% [12.74] & 7.6\% [29.17] & 2.7\% [34.71] & 3.5\% & 0.4\% \\ 
4 & Single & VIPER mock & 6.6\% [13.16] & 5.2\% [12.64] & 5.2\% [24.32] & 0.0\% & 0.0\% & 0.0\% \\ 
5 & Double C1 & D16 & 33.8\% [13.34] & 25.0\% [12.84] & 11.0\% [35.90] & 5.3\% [29.80] & 4.4\% & 2.6\% \\ 
6 & Double C1 & D16 mock & 8.0\% [13.16] & 8.0\% [12.82] & 4.0\% [24.80] & 0.0\% & 0.0\% & 0.0\% \\ 
7 & Double C1 & VIPER & 25.5\% [13.23] & 23.5\% [12.76] & 11.5\% [25.20] & 8.6\% [27.34] & 4.5\% & 4.9\% \\ 
8 & Double C1 & VIPER mock & 11.5\% [13.28] & 9.3\% [12.55] & 4.1\% [26.21] & 2.7\% [38.12] & 0.0\% & 1.3\% \\ 
9 & Double C2 & D16 & 19.9\% [13.32] & 26.3\% [12.78] & 14.0\% [32.70] & 6.6\% [38.40] & 5.1\% & 3.9\% \\ 
10 & Double C2 & D16 mock & 5.6\% [13.18] & 4.0\% [12.71] & 7.2\% [22.50] & 1.3\% [36.40] & 0.0\% & 0.0\% \\ 
11 & Double C2 & VIPER & 26.1\% [13.08] & 29.6\% [12.54] & 12.7\% [30.49] & 11.1\% [27.78] & 3.8\% & 4.9\% \\ 
12 & Double C2 & VIPER mock & 15.5\% [12.98] & 16.0\% [12.99] & 5.4\% [27.36] & 4.0\% [20.10] & 0.7\% & 2.7\% \\ 

   \hline
  \end{tabular}%
  \label{tab:inco}%
\end{table*}%

Rows 5-12 in Table.~\ref{tab:inco} lists the percentage of sample with discrepancy $\Delta b>\sigma_{90b}$ and $\Delta N>\sigma_{90N}$ for double lines. Here the $\sigma_{90b}$ and $\sigma_{90N}$ are taken from Fig.~\ref{fig-result3} for first and second component. Notably, we observe a 33\% sample with $\Delta b>\sigma_{90b}$ for S/N$<20$ compared to \citetalias{Danforth2016} true estimates (see Table.~\ref{tab:inco}). However, this reduces to 8\%  when utilizing ``centered'' mock data (rows 6). This trend of lower inconsistencies for mock data is evident for both components and both studies by \citetalias{Danforth2016}  and VIPER.  Consistent with single absorption lines, predictions for double lines exhibit fewer inconsistencies when $\Delta N > \mathrm{\sigma_{90N}}$ and S/N$>20$. }

As mentioned previously, while plotting the $b$ and  N$_{\rm HI}$, we considered only the common lines. However, in a real-life scenario, we would acquire single and double lines classified by our initial CNN model, which would then serve as input to the respective regression models, yielding distributions of $b$ and N$_{\rm HI}$. Out of the 784 single lines, our classification model correctly identified 684 as single lines and misclassified 58 double lines as single, resulting in a total of 742 single lines. Similarly, our model identified 238 double lines and misclassified 100 single lines as double, resulting in a total of 338 double lines. Consequently, we input 742 single lines into our first regression model, resulting in 742 sets of $b$ and N$_{\rm HI}$ values. Similarly, we input 338 double lines, as identified by our classification model, into our second regression model, resulting in 338$\times2$ values of $b$ and N$_{\rm HI}$ values. In Fig.~\ref{fig-result2_real3}, we depict this distribution of $b$ and N$_{\rm HI}$ in comparison to \citetalias{Danforth2016} and VIPER. We observe that the distribution of column density remains comparable. However, the Doppler width values are affected by three issues: firstly, 30\% of double lines are contaminated by single lines, and secondly, the parameter estimation for double lines is less accurate, and lastly, the centering of the lines is not performed for double lines.

Overall, our analysis reveals that while our algorithm performs reasonably well for single lines, it falls short of the performance achieved by \citetalias{Danforth2016} and VIPER in accurately reproducing the $b$ and N$_{\rm HI}$ values for double lines. This discrepancy highlights that part of the issue lies within our classification algorithm, which does not perform as anticipated. To further investigate, we tested our regression algorithms on the mock datasets we had prepared. In these tests, as highlighted in the previous section, the classification algorithm shows performance in line with expectations. We also find from the mock analysis that the centering of the absorption lines plays a vital role in calculating the physical parameters using the regression algorithm. Moreover, the predictions are found to be more consistent for S/N$>20$.

The successful application of regression analysis to both real observed data and mock data, in conjunction with the prior classification of absorption lines, validates the effectiveness of our comprehensive approach. Our study demonstrates that the ML algorithm yields statistically comparable results to traditional fitting methods. Furthermore, its minimal computational time renders it highly advantageous for large datasets. For instance, while manually fitting a Voigt profile takes at least 1-2 minutes and semiautomated codes like VIPER require 1-2 seconds, the ML algorithm can provide this information in just 0.0002 seconds. This efficiency underscores the potential of ML in efficiently managing complex absorption line analysis, especially considering the substantial demands of data processing.

\section{Main results and discussion}
\label{sect-disscuss}
The low-redshift ($z < 1$) \lya forest is crucial for understanding the evolution of the IGM, galaxy formation, and unresolved baryon fractions. Despite its significant potential, extracting information from the \lya forest presents considerable challenges, especially when using Voigt profile fitting for its numerous absorption lines. Therefore, developing an ML algorithm for Voigt profile fitting is essential for the success of upcoming large astronomical surveys.

In this study, we developed a two-part ML algorithm, FLAME, using CNNs to identify the number of Voigt profiles that best fit a given \lya absorption system and then determine the $b$ and N$_{\rm HI}$ for each profile (see Fig.~\ref{fig-flowchart}). We trained these CNNs with approximately $10^6$ low-$z$ Voigt profiles, synthesized to mimic real data from HST-COS. Given that the majority (96\%) of the \lya lines in the existing high S/N HST-COS data can be fitted with single or double-line profiles (see \citetalias{Danforth2016} and \citetalias{Prakash2017}), we designed our first ML algorithm to classify the lines as either single or double profiles. The second stage of our ML algorithm comprises two networks: the first targets single-line profiles to determine $b$ and N$_{\rm HI}$, while the second is tailored for double-line profiles, determining $b$ and N$_{\rm HI}$ for both components, as well as their velocity separation.

Evaluating the algorithms on simulated \lya lines showcases its impressive performance. The classification algorithm (Fig.~\ref{fig-classify_network}) correctly identifies $\geq 98$\% of the single absorptions and $\geq 90$\% of the double lines (Fig.~\ref{fig-confusion_matrix}). The regression algorithms in the second stage for single Voigt profiles determine values of $b$ and N$_{\rm HI}$ robustly. 
For 68\% of the single lines, the predicted $b$ values lie within within $^{+2.29}_{-1.06}$ km s$^{-1}$, and  log N$_{\rm HI}$  within $^{+0.18}_{-0.16}$ cm$^{-2}$ (Fig.~\ref{fig-result2}). Whereas for 68\% of double lines, the regression algorithm predicts,  $b$ within $^{+3.80}_{-3.82}$ km s$^{-1}$, log $N_{\rm H I}$  within  $^{+0.35}_{-0.43}$ cm$^{-2}$ and velocity separation of both lines $\Delta v$ within $^{+10.42}_{-10.43}$ km s$^{-1}$ of the true values (Fig.~\ref{fig-result3}).
The model demonstrates a close match between predicted and actual parameters, confirming its accuracy. The minimal scatter and negligible bias in predictions underscore its reliability across a broad range of parameters. Nonetheless, the model shows limitations with nearly saturated lines, data with high noise levels, or significant optical depth contrasts, suggesting areas for further enhancement.

We evaluated the performance of our ML algorithms across different parameters, including S/N,  $b$-parameter, N$_{\rm HI}$, and the velocity difference ($\Delta v$) between two absorption lines. Fig.~\ref{accuray_snr} shows the classification accuracy trends with respect to these parameters. As expected, accuracies are notably lower for small values of S/N in simulated datasets. However, excluding cases with these parameters in the lower percentile (S/N$>20$) consistently yields accurate accuracy above 94.2\%, which is promising. Though in this analysis, we test for the wide range of parameters, however, if we established thresholds for these parameters, indicating conditions under which the model provides accurate estimates for $b$, N$_{\rm HI}$, and S/N above 97.5\%. The identified thresholds are as follows: S/N $>$ 20, $b$-parameter $<$ 40 \kms and N$_{\rm HI}$ $<$ 14 cm$^{-2}$.

We applied the algorithm to HST-COS data, focusing on a selected subset of 1,400 absorption lines, and evaluated its performance against two methods of Voigt profile fitting: one using the fits provided by \citetalias{Danforth2016} and the other using the automated Voigt profile fitting code VIPER \citepalias{Prakash2017}. We observed that the accuracy of our classification algorithm decreased by approximately 10\% compared to the simulated dataset. Specifically, we achieved an accuracy of 80\% in comparison with the fits by \citetalias{Danforth2016} and 85\% when compared with the fits from VIPER. Nevertheless, the regression algorithms showed reasonably good agreement in predicting $b$ and N$_{\rm HI}$ values, closely matching their distributions from both VIPER and \citetalias{Danforth2016}. Any discrepancies in the distribution mainly arise from the inherent difficulty in emulating the real observations.

In all our trials, we observed that the real dataset exhibits reduced accuracy compared to the simulated dataset. To understand the reasons behind the lower accuracy of our algorithms on real data, we generated mock data mirroring the parameters of real data fits from both \citetalias{Danforth2016} and VIPER. Applying our algorithm to this mock data, we found that the accuracy remains stable, showcasing an agreement between the predicted number of lines in the profiles and the values of $b$ and N$_{\rm HI}$. This finding suggests that the differences in accuracies between simulated and real datasets can be to some extent ascribed to the inherent complexities in modeling real data, which often contain nuances difficult to replicate accurately in simulations.

The simulated data may fall short of capturing the complexities, variations, and noise inherent in the real data. For example, accurately replicating the inherent noise patterns and instrument-specific calibration errors in simulations is extremely challenging. Strategies such as incorporating a fraction of real data into the training set may not yield the desired effectiveness in this case, as the real data sample size is much lower compared to the size of the simulated data used in this analysis. The ongoing and upcoming large-scale spectroscopic surveys, such as DESI, 4MOST, WEAVE, and PFS, are designed to enhance the availability of real spectra significantly. This influx of real data is expected to play a crucial role in enabling ML models to improve their performance further at higher redshifts. However, currently for lower redshifts, we do not have such large-scale spectroscopic surveys and hence rely on simulated data.

Our study highlights the effectiveness of ML in analyzing \lya absorption lines in quasar spectra, showcasing capability for both classifying the number of lines and estimating physical parameters. Through evaluations with real and mock data, we show that ML not only matches the accuracy of traditional methods but also significantly reduces computational effort, proving especially beneficial for large datasets where traditional approaches falter due to time and labor demands.

\section{Summary}
\label{sect-summary}
We present FLAME, a two-part ML algorithm to fit low-$z$ \lya absorption lines using CNNs. This algorithm effectively determines the optimal number of Voigt profiles for \lya absorption systems and determines the Doppler parameter $b$ and neutral hydrogen column density N$_{\rm HI}$ for each profile. 
FLAME shows impressive accuracy, correctly identifying over 98\% of single absorptions and over 90\% of double lines in simulated data that mimic the real \lya~absorption lines from HST-COS. 
For 90\% of single lines, the FLAME accurately predicts $b$ values within 
$\approx \pm{8}$ km s$^{-1}$ and log N$_{\rm HI}/ {\rm cm}^2$ values within $\approx \pm 0.3$. 
For double lines, it shows slightly lower accuracy, predicting $b$ within $\approx \pm 15$ km s$^{-1}$ and log N$_{\rm HI}/{\rm cm}^2$ within $\approx \pm 0.8$. 

Despite its impressive performance on simulated data, FLAME's accuracy in classifying lines  decreases  by about 10\% when applied to real HST-COS data. Nevertheless, there is good agreement between the predicted $b$ and N$_{\rm HI}$ distributions with other Voigt profile-fitting algorithms, such as VIPER (\citetalias{Prakash2017}), and the fits from \citetalias{Danforth2016} for single lines. However, the fits need to be improved for double lines.

Our analysis with mock HST-COS data, crafted to reflect the parameters of real data, shows that FLAME can maintain stable accuracy, mirroring its success with simulated datasets. This underscores that the primary discrepancies in accuracy when applied to simulated versus real data stem from the challenges in fully capturing the complexities of real data in the simulated training dataset. Despite these hurdles,  FLAME successfully demonstrates the feasibility of employing ML to fit Voigt profiles, highlighting the potential for ML in analyzing absorption lines. 

Moving forward, we aim to refine FLAME, particularly to improve its performance with real data. Our efforts will include examining the specific challenges posed by real data and assessing how these affect accuracy. Including additional spectral information, such as metal lines, similar to the approach taken by \citetalias{Danforth2016}, appears to be a promising method for enhancing accuracy. Additionally, incorporating real data into our training samples is a key strategy we believe will help align FLAME more closely with practical applications. The upcoming influx of data from spectroscopic surveys like DESI, 4MOST, WEAVE, and PFS is expected to benefit ML models significantly by providing richer training datasets. This effort will lead to more precise characterizations of the \lya forest by future developments of FLAME.

\begin{acknowledgements}

We thank our anonymous referee for their positive feedback and useful comments. PJ acknowledges Dr. hab Maciej Bilicki for useful discussion. The Polish National Science Center supported PJ through grant no. 2020/38/E/ST9/00395. VK is supported through the INSPIRE Faculty Award (No. DST/INSPIRE/04/2019/001580) of the Department of Science and Technology (DST), India, and by NASA through grant number HST-AR-17048.003 from the Space Telescope Science Institute, which is operated by the Associated Universities for Research in Astronomy, Inc., under NASA contract NAS 5-26555. 
A partial finances during the manuscript are supported by the Young Scientist Award 2023, won by PJ. PJ acknowledges the CFT and ARIES computer facility to provide high-performance computers. MV acknowledges support from DST-SERB in the form of a core research grant (CRG/2020/1657). The authors acknowledge the assistance of tools such as Grammarly and ChatGpt in polishing the text at various places.\end{acknowledgements}

\bibliography{ref}{}
\bibliographystyle{aa}
\end{document}